\documentclass[12pt]{article}
\pdfoutput=1
\usepackage{graphicx,epstopdf,amssymb,amsfonts,amsmath,amsthm,array,
mathrsfs,amscd}
\usepackage{hyperref}
\newcommand{\pbs}[1]{\let\temp=\\#1\let\\=\temp}
\numberwithin{equation}{section}
\def\be{\begin{equation}}\def\ee{\end{equation}}
\def\cvp{\raise 2pt\hbox{,}} 
 \def\tr{\mathop{\rm tr}\nolimits}

 \def\d{{\rm d}} 
 \def\uN{\text{U}(N)}

\def\la{\lambda}\def\La{\Lambda}
\def\vol{\mathop{\text{vol}}\nolimits}
\def\stot{S_{\text{tot}}}
\def\gh{\mathop{\text{gh}}\nolimits}
\def\lieh{\mathfrak h}\def\lieg{\mathfrak g}
\def\lieq{\mathfrak g/\mathfrak h}
\def\A{\mathscr A}\def\AW{\mathscr A_{\text W}}\def\L{\mathscr L}\def\Int{\mathscr I}\def\dW{\d_{\text W}}
\def\ABRST{\mathscr A_{\text{BRST}}}
\def\Ainv{\mathscr A_{\text{inv}}}\def\Abas{\mathscr A_{\text{bas}}}
\def\AC{\mathscr A_{\text C}}\def\dC{\d}
\def\dCh{\delta}
\def\ACh{\mathsf A_{\text C}}
\def\AWeil{\mathsf A_{\text{Weil}}}
\theoremstyle{plain}

\theoremstyle{definition}
\theoremstyle{remark}

\def\plb#1#2#3{{\it Phys.\ Lett.\ }{\bf B #1} (#2) #3}
\def\npb#1#2#3{{\it Nucl.\ Phys.\ }{\bf B #1} (#2) #3}
\def\npps#1#2#3{{\it Nucl.\ Phys.\ Proc.\ Suppl.\ }{\bf #1} (#2) #3}

\def\prd#1#2#3{{\it Phys.\ Rev.\ }{\bf D #1} (#2) #3}

\def\cmp#1#2#3{{\it Comm.\ Math.\ Phys.\ }{\bf #1} (#2) #3}

\def\ap#1#2#3{{\it Ann.\ of Phys.\ }{\bf #1} (#2) #3}

\def\ptp#1#2#3{{\it Prog.\ Theor. Phys. }{\bf #1} (#2) #3}

\def\imath#1#2#3{{\it Invent math }{\bf #1} (#2) #3}

\begin{document}
%
%
{\pagestyle{empty}
\parskip 0in
\

\vfill
\begin{center}



{\LARGE Partial Gauge Fixing and Equivariant Cohomology}

\vspace{0.4in}

Frank F{\scshape errari}
\\
\medskip
{\it Service de Physique Th\'eorique et Math\'ematique\\
Universit\'e Libre de Bruxelles and International Solvay Institutes\\
Campus de la Plaine, CP 231, B-1050 Bruxelles, Belgique}
\smallskip
{\tt frank.ferrari@ulb.ac.be}
\end{center}
\vfill\noindent

Given a gauge theory with gauge group $G$, it is sometimes useful to find an equivalent formulation in terms of a non-trivial gauge subgroup $H\subset G$. This amounts to fixing the gauge partially from $G$ down to $H$. We study this problem systematically, both from the algebraic and the path integral points of view. We find that the usual BRST cohomology must be replaced by an equivariant version and that the ghost Lagrangian must always include quartic ghost terms, even at tree level. Both the Cartan and Weil models for equivariant cohomology play a r\^ole and find natural interpretations within the physics framework. Applications include the construction of D-brane models of emergent space, the 't~Hooft Abelian projection scenario in quantum chromodynamics and the formulation of the low energy effective theories of grand unified models.

\vfill

\medskip
%
\begin{flushleft}
\today
\end{flushleft}
\newpage\pagestyle{plain}
\baselineskip 16pt
\setcounter{footnote}{0}

}

\section{\label{s1} Introduction}
\subsection{General presentation}\label{genpresSec}

It is well known that the so-called gauge symmetry is a misnomer. A gauge symmetry is really a redundancy in the description of the model under consideration and thus should better be referred to as a ``gauge redundancy.'' This redundancy is usually introduced because it can yield great simplifications in the formulation of the model. For example, in the theory of the fundamental interactions, the gauge theoretic formulation provides a local and manifestly Lorentz invariant description of the physics. However, to extract the physical information from the theory, the gauge redundancy must be waived. This is done through a gauge-fixing procedure which, morally speaking, amounts to picking a unique representative in field space on each orbit of the gauge group. For example, in a gauge field theory with  gauge group $G$ and fields $\Phi$, we can impose conditions
\be\label{gf1} F_{A}(\Phi) = 0\, ,\quad 1\leq A\leq\dim G\, ,\ee
chosen in such a way that there is one and only one solution on each orbit of $G$. More generally, the gauge-fixing procedure can be done within the BRST framework \cite{BRST}. We enlarge the set of fields by adding so-called ghost fields, anti-ghost fields, ``Lagrange multiplier'' fields, possibly ghosts for ghosts, etc., denoted collectively by $\tilde\Phi$. The field algebra generated by both $\Phi$ and $\tilde\Phi$ is endowed with a grading, the ghost number, and a graded differential, the nilpotent BRST operator $s$. The gauge fixing is done by adding to the original gauge invariant action $S(\Phi)$ a gauge-fixing functional $-s\Psi$, where the gauge-fixing fermion $\Psi(\Phi,\tilde\Phi)$ is of ghost number $\gh\Psi=-1$. The total action is thus
\be\label{stotal} \stot(\Phi,\tilde\Phi) = S(\Phi) - s\Psi (\Phi,\tilde\Phi)\, .\ee
Of course, by construction, the total action is not gauge invariant. Nevertheless, the gauge invariance of the original action implies that the expectation value of any gauge invariant operator will be independent of the gauge-fixing fermion $\Psi$, within a broad class of possible choices yielding consistent gauge-fixing terms. This is equivalent to saying that the physics is encoded in the cohomology of the BRST operator $s$ at ghost number zero. Even more generally, the gauge fixing can be done within the BV formalism \cite{BV}. This is most useful in particular in the case of open gauge algebras. Even though the problem we consider in the present paper does share similarities with the problem of open gauge algebras, as we shall briefly mention later, the BRST framework, suitably generalized, will be perfectly appropriate for our purposes. We shall also assume that the gauge theory we start with has a closed irreducible gauge algebra, since the main applications we have in mind are in the context of Yang-Mills gauge theories. The case of closed reducible algebras could certainly be treated with only simple technical modifications; the case of open algebras would require us to start with the BV framework.

The problem we are going to focus on is a generalization of the standard gauge-fixing procedure that we have just briefly reviewed, for which the gauge symmetry is only partially fixed down to a non-trivial subgroup $H\subset G$. We shall see that there are in general many ways to do this, the ambiguity in the procedure being governed by an equivariant version of the BRST cohomology. All the resulting models are physically equivalent gauge theories with gauge group $H$. Upon gauge-fixing $H$, they yield the same partition function and gauge-invariant correlators as the original gauge theory with gauge group $G$. 

Amongst the possible applications of the partial gauge fixing procedure, we have in mind especially three of them, which are related to our recent work \cite{ferfund}. 

\noindent\emph{Low energy effective actions in grand unified models:} in this case one considers a gauge theory with gauge group $G$ in a Higgs phase. The low energy physics is then described in general by a gauge theory with ``unbroken'' gauge group $H\subset G$. To derive and study this low energy gauge theory, it is very natural to work in a framework where the original gauge symmetry is partially fixed down to $H$. This point of view was  advocated in \cite{weinberg}.

\noindent\emph{Abelian projection:} as argued by 't~Hooft in \cite{tHooftAP}, the monopole condensation picture for confinement in QCD may be most naturally understood when the theory is reformulated as an Abelian gauge theory. This amounts to partially fixing the gauge from $\text{SU}(3)$ down to the maximal Abelian subgroup $\text{U}(1)^{2}$, or more generally from $\text{SU}(N)$ down to $\text{U}(1)^{N-1}$. In spite of the enormous literature on this theme, a full analysis of this partial gauge fixing does not seem to have appeared before. Our results clarify some of the interesting literature on this subject.

\noindent\emph{D-brane models}: on a stack of $N$ D-branes in string theory lives a $\uN$ gauge theory. It is sometimes useful to separate this stack of branes into several sets, for example into two sets of $N_{1}$ and $N_{2}$ branes. The natural description of the physics is then in terms of a $\text{U}(N_{1})\times\text{U}(N_{2})$ gauge group, which amounts to a partial fixing of the original gauge symmetry. This point of view is particularly interesting when $N_{1}\gg N_{2}$. In this limit, one can in principle replace the $N_{1}$ branes with the emerging holographic geometry sourced by the original gauge theory. The effective action for the $N_{2}$ branes should then coincide with a probe brane action for branes moving in this emerging geometry! This is the point of view advocated in \cite{fer1,ferfund}. Moreover, the emerging geometry can in principle be read off from the effective action \cite{fer1,ferall,ferfund}. This effective action is a $\text{U}(N_{2})$ gauge theory, obtained by the partial gauge-fixing of the original $\text{U}(N_{1}+N_{2})$ gauge symmetry. 

\noindent This last, most modern application was the main motivation to start the present investigations. The gauge-dependence of the partially gauged-fixed effective action seems to be related to the problem of bulk locality \cite{ferfund}. Actually, the D-brane picture reveals unexpected deep links between D-brane physics, holography, the Abelian projection scenario, bulk locality and the low energy physics of gauge theories with a broken gauge symmetry, as in grand unified models; eventually, these ideas lead to a new approach to gauge theories, see \cite{ferfund} for more details.

\medskip

\noindent\emph{Short comments on the existing literature}

In view of the many possible applications, in particular to the classic problems of grand unification and the Abelian projection, to which a huge literature is devoted, it may seem quite surprising that a full discussion of the partial gauge-fixing procedure, from first principles, has never appeared before. However, in spite of a rather extensive search, we have not been able to find any. The original reference, in which the problem is very clearly stated, is \cite{weinberg}. However, very few details are given in this work and, unfortunately, the ansatz which is proposed for the ghost Lagrangian is not correct. Moreover, the BRST symmetry structure was not investigated at all in \cite{weinberg}. The many works that subsequently refer to \cite{weinberg} seem to believe that the partial gauge-fixing procedure studied in this reference is the same as a background field gauge. The background field gauge and the partial gauge-fixing are actually two completely distinct concepts and the background field gauge is not discussed at all in \cite{weinberg} or in our present work.\footnote{In the background field gauge approach, \emph{the gauge symmetry $G$ is completely fixed}, but with a particular gauge choice which depends on a classical background gauge field, chosen in such a way that a new gauge symmetry appears in the problem. In the partial gauge-fixing procedure, \emph{the gauge symmetry is only partially fixed from $G$ down to a non-trivial subgroup $H$}. The unbroken gauge symmetry $H$ is then part of the original gauge group of the model and has nothing to do with a classical background gauge field symmetry. This crucial difference implies in particular that the structure of the ghost Lagrangian is completely different in the two cases.} In the literature on the Abelian projection scenario, which mainly focuses on the $\text{SU}(2)\rightarrow\text{U}(1)$ partial breaking,\footnote{We shall see below that many simplifications occur in this case, mainly because $\text{SU}(2)/\text{U}(1)$ is a symmetric space.} very interesting remarks have appeared about what could be the most appropriate ghost Lagrangian to use. In particular, it was noticed in \cite{Kondo} that the introduction of a particular quartic ghost term could greatly simplify the renormalization properties of the theory, see also \cite{renAG}. Such a quartic ghost coupling in the $\text{SU}(2)$ model was also considered in \cite{Schaden}, with a point of view and motivations coming from lattice gauge theories, see \cite{Shamir} for very relevant further developments. An equivariant BRST differential was also built in \cite{Schaden}, which is a special (and truncated) case of the differential $\dCh$ that we shall introduce in our work. These papers on the Abelian projection and on the lattice can thus be rightly regarded as precursors of some of our results which, in return, hopefully illuminate and provide a full justification of the prescriptions used in \cite{Kondo,Schaden,Shamir}. 

However, the signification of the quartic ghost terms that emerges from our work is very different from the one discussed in the Abelian projection literature (but is consistent with \cite{Schaden, Shamir}). It turns out that the quartic ghost couplings play a central r\^ole in any partial gauge-fixing procedure $G\rightarrow H$ and must actually be present in the \emph{tree-level} ghost Lagrangian. \emph{This is totally independent of any renormalization consideration.} Instead, we shall see that \emph{the quartic ghost couplings are required by gauge invariance}, which would be violated if these terms were omitted! A very neat example of this phenomenon will be presented in \cite{MMbranes} in the context of the zero-dimensional one matrix model, where obviously renormalization is not an issue. Our result about the quartic ghost terms is thus totally unrelated to the usual quartic ghost couplings that are added to renormalize the (non gauge-invariant) 1PI effective action when non-linear gauge-fixing conditions are used. In this latter traditional and well-known case, the quartic ghost terms are BRST-exact terms that do not contribute to the gauge-invariant observables.

\subsection{Statement of the problem and its solution}\label{problemsec}

Let us now state precisely the problem we want to solve and briefly explain the solution we shall find. We work in Euclidean signature. Going to the Minkowskian signature is trivial and essentially amounts to replacing the factors $e^{-S}$ in the path integrals with $e^{iS}$.

Let $Z$ be the partition function for a gauge theory (think of a Yang-Mills model), with gauge group $G$, defined by the path integral
\be\label{Zdef} Z = \int\! D\Phi D\omega_{A}D\bar\omega_{A}D\bar\la_{A}\, e^{-S(\Phi) + s_{G}\Psi_{G}(\Phi,\omega,\bar\omega,\bar\la)}\, ,\ee
where $1\leq A\leq \dim G$ and $\omega_{A}$, $\bar\omega_{A}$ and $\bar\la_{A}$ are the ghosts, antighosts and Lagrange multiplier fields in the adjoint representation of $G$ respectively. 
The gauge invariant action $S(\Phi)$ may include sources for arbitrary gauge invariant operators. We can thus think of $Z$ as a generating functional encoding all the information about the theory. The gauge-fixing fermion $\Psi_{G}$ completely fixes the gauge for $G$ in the usual way. The differential $s_{G}$ is the standard BRST operator for the gauge group $G$. 

Let $H\subset G$ be a non-trivial subgroup of $G$. We split the set of fields $\Phi$ into two subsets,
\be\label{phiPhisep} \{\Phi\} = \{\varphi,\phi\}\, ,\ee
such that $\varphi$ and $\phi$ each belong to a representation of $H$ (i.e.\ $\varphi$ and $\phi$ do not mix under a gauge transformation belonging to $H$). This is the only condition we impose on the splitting, which may otherwise be completely arbitrary and in particular is not unique. In the application to the low energy effective actions in grand unified models, as in \cite{weinberg}, the fields $\varphi$ are always chosen to be the low mass fields and $\phi$ the unification-scale fields, but we emphasize that this distinction according to mass is not necessary to develop the formalism and does not make sense a priori in the other applications mentioned above.

Our goal is to find actions $S_{H}(\varphi)$, which are gauge invariant under $H$ and thus define gauge theories with gauge group $H$, which all have precisely the same partition function $Z$ \eqref{Zdef} as the original gauge theory with gauge group $G$. In other words, we want 
\be\label{ZHdef} Z = \int\! D\varphi D\omega_{a}D\bar\omega_{a}D\bar\la_{a}\, e^{-S_{H}(\varphi) + s_{H}\psi_{H}(\varphi,\omega,\bar\omega,\bar\la)}\, ,\ee
where $1\leq a\leq \dim H$ and $\omega_{a}$, $\bar\omega_{a}$ and $\bar\la_{a}$ are the ghosts, antighosts and Lagrange multiplier fields in the adjoint representation of $H$ respectively. The gauge-fixing fermion $\psi_{H}$ completely fixes the gauge for $H$ in the standard way, $s_{H}$ being the usual BRST differential for the gauge group $H$.

We shall show that the possible (non-local) actions $S_{H}$ are given by a path integral formula of the form
\be\label{SHdef} e^{-S_{H}(\varphi;\psi_{G/H})} = \int\! D\phi D\Omega_{i}D\bar\Omega_{i}D\bar\La_{i}\,
e^{-S(\varphi,\phi) + \delta\psi_{G/H}(\varphi,\phi,\Omega,\bar\Omega,\bar\La)}\, ,\ee
where $1\leq i\leq\dim G - \dim H$ and $\Omega_{i}$, $\bar\Omega_{i}$ and $\bar\La_{i}$ are ghosts, antighosts and Lagrange multiplier fields that are required to gauge fix $G$ down to $H$. The gauge-fixing fermion $\psi_{G/H}$ is a functional of ghost number $\gh\psi_{G/H}=-1$ which must be gauge invariant under $H$. The operator $\delta$ is a graded derivation which is \emph{not} nilpotent, but whose square is a gauge transformation belonging to $H$ (one says that $\delta$ is an equivariant differential with respect to $H$) and which commutes with the $H$ gauge transformations.

The actions $S_{H}$ defined by \eqref{SHdef} are manifestly gauge invariant under $H$. However, they \emph{do} depend on the choice of the gauge-fixing fermion $\psi_{G/H}$. The main point is that the full partition function defined by the path integral \eqref{ZHdef} does \emph{not} depend on $\psi_{G/H}$ (nor of course on $\psi_{H}$). All the actions $S_{H}$, parameterized by the gauge-fixing fermion $\psi_{G/H}$, are thus all physically equivalent to the original action $S$ invariant under $G$. 

The most remarkable feature of the ghost Lagrangian produced by computing 
$\delta\psi_{G/H}$ is that it always includes quartic ghost terms. \emph{This is a tree-level effect and is required by consistency}. Indeed, the remarkable property that the partition function \eqref{ZHdef} does not depend on $\psi_{G/H}$, even though $S_{H}$ itself strongly depends on it, relies crucially on the presence of the quartic ghost terms in the partially gauge-fixed action. In other words, the quartic ghost terms are required by gauge invariance. A very instructive example showing explicitly how this works is presented in \cite{MMbranes}. This discussion also makes very clear a point that we have already emphasized in the previous subsection: the tree-level quartic ghost terms that we find in the partial gauge-fixing procedure are of a completely different nature than the well-known quartic ghost counterterms that one has to include in loop calculations to define the renormalized (and non gauge invariant) 1PI effective action when using a non-linear full gauge-fixing condition. These latter traditional terms do not contribute to gauge-invariant quantities.

Let us note that there is one very special case where we can actually do without the quartic ghost terms. This case corresponds to non-renormalizable strict partial gauge-fixing condition \`a la Landau, imposed via a $\delta$-function in the path integral. This is very inconvenient. For practical calculations, one usually imposes the gauge-fixing with a Gaussian weight, which then must necessarily come with quartic ghost interactions in the tree-level ghost Lagrangian.  

\subsection{Plan of the paper}

There are two different routes that one can follow in order to study the partial gauge-fixing procedure and in particular to derive Eq.\ \eqref{SHdef}. 

The first route uses an abstract algebraic point of view, based on equivariant cohomology. This route delivers right away the full structure and deepest understanding of the solution. The advantage of this approach is that it is completely straightforward and natural. This is how the results were originally derived by the author. Its drawback is that one must be familiar with the Cartan and Weil models of equivariant cohomology before starting.

The second route is completely elementary. It starts from \eqref{Zdef} and goes to \eqref{ZHdef} and \eqref{SHdef} through a series of simple steps. This is undoubtedly the most straightforward and shortest route if one is only interested in the solution and on practical calculations using this solution, being ready to give up a full, deeper understanding. The main drawback of this route is that it uses some cunning tricks which seem to work only by miracle. In the abstract algebraic approach, these tricks are automatically implemented.

We shall present both routes in the following, devoting Sections 2 and 3 to the algebraic point of view and Section 4 to the elementary approach. To make the paper self-contained, we have included in Section 2 an introduction to the standard algebraic notions underlying the BRST framework and equivariant cohomology. No prior knowledge of the standard models (Cartan or Weil) of equivariant cohomology is assumed in our presentation. The only original material in this Section seems to be the notion of equivariant trivial pairs. In Section 3, we show that the ansatz \eqref{SHdef} follows immediately from the Cartan model of equivariant cohomology, which can also be interpreted in the present physics context as a ``ghosts for ghosts'' approach to the problem of partial gauge-fixing. We then prove, using the Weil model, that the partition function \eqref{ZHdef} does not depend on the choice of $\psi_{G/H}$ and coincides with the partition function \eqref{Zdef} of the original theory. In Section 4, we start our analysis anew and rederive the results in a completely elementary way. This includes a path integral analysis \`a la Faddeev-Popov using a trick due to Zinn-Justin \cite{ZJtrick}. We have tried to make this section independent of the material presented after Section \ref{gtSec}, so that the reader not interested in the algebraic point of view may jump directly from the end of \ref{gtSec} to Section \ref{s4}. In particular, all the references we make to the rest of Section \ref{s2} and to Section \ref{s3} are helpful to put into perspective the tricks used in the computations but are not required to follow the derivations. In Section \ref{s5}, we present the detailed explicit formulas for a general ``Yang-Mills D-brane system,'' which corresponds to the partial breaking $\uN\rightarrow\text{U}(N_{1})\times\cdots\times\text{U}(N_{r})$ with $\sum_{I}N_{I} = N$. The usual Abelian projection is found when $r=N$ and all the $N_{I}=1$, the case most studied in the literature being $N=2$, $N_{1}=N_{2}=1$. Finally, we conclude in Section \ref{s6}, providing in particular a window on the use and interpretation of our results in the context of emergent space models \cite{ferfund}.

\noindent\emph{Remark on notations}: most of the time, we shall use a ``super-index'' notation, for which the indices actually include both spacetime and group indices. For example, a variable $\bar\Omega_{i}$ is really a field $\bar\Omega_{i}(x)$ and a sum over $i$ like $\bar\Omega_{i}\bar\La_{i}$ is really $\int\!\d x\, \bar\Omega_{i}(x)\bar\La_{i}(x)$.

\section{\label{s2} Algebraic preliminaries}

In this section, we introduce the algebraic framework in which the partial gauge-fixing procedure can be naturally explained and understood. Most of this material is of course not new. If not for the notion of equivariant trivial pairs presented in \ref{cohosec}, which does not seem to have been discussed before, the only claim of originality may be in the style of presentation which is self-contained and adapted to the problem we want to tackle. For more details, we refer the reader to \cite{algebra} and references therein.

\subsection{Group theory}\label{gtSec}

We consider a gauge group $G$ which is compact and semi-simple (the inclusion of $\text{U}(1)$ factors, like in the case of $\uN$, is straightforward). We denote by $(\tau_{A})_{1\leq A\leq \dim G}$ a basis of the Lie algebra $\mathfrak g$ in which the Killing form is the unit matrix. In particular, $\lieg$ and its dual $\lieg^{*}$ are identified. The completely antisymmetric structure constants $f_{ABC}$ are such that
\be\label{comrel} [\tau_{A},\tau_{B}] = f_{ABC}\tau_{C}\ee
and satisfy the Jacobi identity
\be\label{Jacobi} f_{E[AB}f_{C]DE} = 0\, .\ee
Let $H$ be a subgroup of $G$. The basis $(\tau_{A})$ of $\mathfrak g$ is chosen in such a way that $(\tau_{a})_{1\leq a\leq\dim H}$ is a basis of the Lie algebra $\mathfrak h$ of $H$. The other, ``broken,'' generators are denoted as $(\tau_{i})_{\dim H+1\leq i\leq\dim G}$ and generate $\mathfrak g/\mathfrak h$. We have
\be\label{Liedec} \lieg = \lieh\oplus\lieq\, .\ee
The indices $A,B,C,$ etc., will always correspond to Lie algebra indices for $G$, the ``unbroken'' indices $a,b,c,$ etc., to Lie algebra indices for $H$  and the ``broken'' indices $i,j,k,$ etc., to Lie algebra indices for the broken generators. When indices of a given type are repeated, a sum is always assumed, as in \eqref{Jacobi}. The commutation relations \eqref{comrel} split as
\begin{align}\label{comab} [\tau_{a},\tau_{b}] &= f_{abc}\tau_{c}\\
\label{comai}
[\tau_{a},\tau_{i}] &= f_{aij}\tau_{j}\\\label{comij}
[\tau_{i},\tau_{j}] & = f_{ija}\tau_{a} + f_{ijk}\tau_{k}\, .
\end{align}
The first relations \eqref{comab} follow from the fact that $H$ is a subgroup, which implies that
\be\label{fabi}f_{abi}=0\, .\ee
The second relations \eqref{comai} follow from $f_{aib}=-f_{abi}=0$. They imply that the $\tau_{i}$ transform in a real representation of $H$ which we denote by $R_{G/H}$. The generators in this representation take the form
\be\label{realrep} (\tau_{a})_{ij}=-f_{aij}\, .\ee
The adjoint representation of $G$ decomposes as
\be\label{Adjdec} \text{Adj}_{G}=\text{Adj}_{H}\oplus R_{G/H}\, ,\ee
mimicking the decomposition \eqref{Liedec}. 

Let us note that two special cases can occur. In the first case, the structure constants $f_{ijk}$ vanish. This yields interesting examples, for which the quotient $G/H$ is a symmetric space. In the second case, the $f_{aij}$'s vanish and thus $\lieg/\lieh$ is itself a Lie algebra. This case is trivial, because the problem of the partial gauge fixing of $G$ down to $H$ is reduced to the standard full gauge fixing of the group $G/H$.

One can easily build group invariants. If $X_{A}$, $Y_{A}$ and $Z_{A}$ are in the adjoint of $G$, then
\be\label{adjGinv} X_{A}Y_{A}=X_{a}Y_{a} + X_{i}Y_{i}\, ,\quad f_{ABC}X_{A}Y_{B}Z_{C}\ee
are $G$-invariant. Similarly, if $X_{a}$, $Y_{a}$ and $Z_{a}$ transform in the adjoint of $H$ and $x_{i}$, $y_{i}$ and $z_{i}$ transform in the representation $R_{G/H}$, then the following combinations
\be\label{Hinv} X_{a}Y_{a}\, ,\quad x_{i}y_{i}\ ,\quad f_{abc}X_{a}Y_{b}Z_{c}\, ,\quad f_{aij}X_{a}y_{i}z_{j}\, ,\quad f_{ijk}x_{i}y_{j}z_{k}\ee
are $H$-invariant. This can be easily checked by using various special cases of the Jacobi identity \eqref{Jacobi}, where the indices $A, B, C, D$ are chosen to be either unbroken or broken indices.

\subsection{$\mathfrak g$-differential algebras}\label{gdiffsec}

\subsubsection*{Basic definitions}

Let us consider a $\mathbb Z$ (or $\mathbb N$) graded superalgebra $\A$,
\be\label{Agrad} \A = \oplus \A_{n}\, .\ee
We denote by $\epsilon$ the $\mathbb Z_{2}$ grading associated with the commuting or anticommuting (even or odd) nature of the variables and by gh the $\mathbb Z$ (or $\mathbb N$) grading. A basic example is the superalgebra of fields, in which case the $\mathbb Z$-grading is the ghost number. Another classic example is the exterior algebra of differential forms over a manifold. The $\mathbb N$-grading is then the degree of the forms.

A \emph{graded derivation $D$ of degree $p$ on $\A$} is a linear map from the spaces $\A_{n}$ to $\A_{n+p}$ such that
\be\label{derdef} D (xy) = (Dx) y \pm x (Dy)\, .\ee
The $\pm$ sign is always a $+$ in the case of even graded derivations and is $(-1)^{\epsilon_{x}}$ in the case of odd graded derivations. Note that, by \eqref{derdef}, the action of a graded derivation is known on the full algebra once it is known on a generating set. One can easily check that the $\mathbb Z_{2}$-graded commutators of graded derivations are graded derivations. For us, a \emph{differential} is a nilpotent odd graded derivation of degree one.

The action of a group $G$ on $\A$ defines an algebra of even graded derivations of degree zero, denoted by $\mathscr L_{A}$ and called the \emph{Lie derivatives}, such that an infinitesimal transformation reads
\be\label{infactG} \delta x = -\varepsilon_{A}\mathscr L_{A}x\, .\ee
The Lie derivatives automatically satisfy the Lie algebra commutation relations
\be\label{comLie} [\mathscr L_{A},\L_{B}] = f_{ABC}\L_{C}\, .\ee
A typical example is the action of a Lie group on a manifold. The action on the exterior algebra of differential forms is then fixed by the action on the coordinates $x^{\mu}$ and the one-forms $\d x^{\mu}$,
\be\label{actionmanifold} \L_{A}x^{\mu} = \xi_{A}^{\mu}(x)\, ,\quad
\L_{A}\d x^{\mu} = \partial_{\nu}\xi^{\mu}_{A}(x)\d x^{\nu}\, ,\ee
where the $\xi_{A}$ are the vector fields associated with the action of $G$ on the manifold. In this case, $\L_{A}$ coincides with the ordinary Lie derivative with respect to $\xi_{A}$. When $\A$ is the field algebra of a gauge theory, the action of $G$ corresponds to the gauge transformations. Remember that the formulas should then be interpreted in a ``super-index'' notation where the indices actually include both spacetime and group indices. Similarly, the gauge group is always really an infinite dimensional group which can be viewed as the infinite direct product of the finite dimensional Lie group $G$ defined at each spacetime point. For example, a gauge group generator should be denoted as $\tau_{A,x}$, where $x$ is a spacetime point. The action on the gauge potential reads
\be\label{LieonA1} \L_{A,x} A_{\mu B}(y) = \partial_{\mu}\delta(x-y) + \delta(x-y) f_{ABC} A_{\mu C}(y)\, .\ee
It is then more convenient to introduce 
\be\label{epsgaguedef}\L_{\varepsilon} = \int\!\d x\, \varepsilon_{A}(x)\L_{A,x}\, ,\ee
such that
\be\label{LieonA2} \L_{\varepsilon}A_{\mu B}(x) = -\partial_{\mu}\varepsilon_{B}(x) + f_{ABC}\varepsilon_{A}(x)A_{\mu C}(x)\, ,\ee
which is the more traditional form of the gauge transformations. This subtlety with the notations should be kept in mind but will not be mentioned any longer.

A \emph{$\lieg$-differential algebra $(\A,D,\Int_{A},\L_{A})$} is a graded superalgebra $\A$ with an action of $G$ represented by the Lie derivatives $\L_{A}$ and endowed with odd graded derivations $D$ and $\Int_{A}$ of degrees one and minus one respectively, such that
\begin{align}\label{ralg1} &D^{2}=0\, ,\quad [\Int_{A},\Int_{B}] = 0\, ,\quad [\Int_{A},D]=\L_{A}\\
\label{ralg2} &  [\L_{A},\L_{B}] = f_{ABC}\L_{C}\, ,\quad
[\L_{A},\Int_{B}] = f_{ABC}\Int_{C}\, ,\quad [\L_{A},D] =0\, .
\end{align}
The bracket we use is a graded commutator, for example $[\Int_{A},D] = \Int_{A}D + D\Int_{A}$ whereas $[\L_{A},\L_{B}]=\L_{A}\L_{B}-\L_{B}\L_{A}$. The first relation in \eqref{ralg1} says that $D$ is a differential. The third relation is known as the Cartan equation. The first relation in \eqref{ralg2} is a consequence of the fact that $G$ acts on $\A$. The third relation can actually be deduced from $D^{2}=0$ and the Cartan equation and thus could be omitted from the definition. 

The structure of $\lieg$-differential algebra will be ubiquitous in the following. The most classic example is again the exterior algebra of differential forms on a manifold, $D$ being the exterior derivative and $\Int_{A}$ the interior products with respect to the vector fields $\xi_{A}$. We shall see several other examples below.

If $(\A,D,\Int_{A},\L_{A})$ is a $\lieg$-differential algebra and $H\subset G$ is a subgroup of $G$, then obviously $(\A,D,\Int_{a},\L_{a})$ is an $\lieh$-differential algebra. If $(\A^{(1)},D^{(1)},\Int_{A}^{(1)},\L_{A}^{(1)})$ and $(\A^{(2)},D^{(2)},\Int_{A}^{(2)},\L_{A}^{(2)})$ are $\lieg$-differential algebras, then $(\A^{(1)}\otimes\A^{(2)},D,\Int_{A},\L_{A})$ is a $\lieg$-differential algebra, with
\begin{align}\label{dsum} D &= D^{(1)}\otimes\mathbb I^{(2)} + (-1)^{\epsilon^{(1)}}\otimes D^{(2)}\\\label{Isum}
\Int_{A} &= \Int_{A}^{(1)}\otimes\mathbb I^{(1)} + (-1)^{\epsilon^{(1)}}\otimes\Int_{A}^{(2)}\\\label{Lsum}
\L_{A} &= \L_{A}^{(1)}\otimes\mathbb I^{(2)} + \mathbb I^{(1)}\otimes\L_{A}^{(2)}\, .
\end{align}
The identity operators $\mathbb I^{(1)}$ and $\mathbb I^{(2)}$ act on $\A^{(1)}$ and $\A^{(2)}$ respectively; $\epsilon^{(1)}$ is the $\mathbb Z_{2}$-grading of the superalgebra $\A^{(1)}$. The relations \eqref{dsum}--\eqref{Lsum} will often be simply denoted $D = D^{(1)} + D^{(2)}$, etc., keeping in mind the odd nature of the graded derivations $D$ and $\Int_{A}$.

\subsubsection*{Connections and curvatures}

An \emph{algebraic connection}, or \emph{connection} for short, of a $\lieg$-differential algebra is a Lie-algebra valued odd element $\theta_{A}$ such that
\be\label{connexiondef} \Int_{A}\theta_{B} = \delta_{AB}\, ,\quad \L_{A}\theta_{B}=f_{ABC}\theta_{C}\, .\ee
These equations translate algebraically the basic properties of a connection on a principal bundle: the first condition says that $\theta$ projects on the vertical subspaces, which are generated by the action of $G$, along the horizontal subspaces; the second condition, which tells that the connection transforms in the adjoint representation, ensures the compatibility of the choice of horizontal subspaces with the group action. In the present algebraic framework, no geometrical interpretation of the relations \eqref{connexiondef} are needed; they are simply abstract algebraic requirements. In the same spirit, we can define the algebraic curvature of the connection $\theta$ by the equation
\be\label{curvdef} F_{A} = D\theta_{A} + \frac{1}{2}f_{ABC}\theta_{B}\theta_{C}\, .\ee
Using the axioms of $\lieg$-differential algebras, in particular the Cartan equation, one then shows that
\be\label{curveq} \Int_{A}F_{B}= 0\, ,\quad \L_{A}F_{B}=f_{ABC}F_{C}\, ,\quad D F_{A} = f_{ABC}F_{B}\theta_{C}\, .\ee
We see that the curvature automatically transforms in the adjoint representation and satisfies the Bianchi identity.

\subsubsection*{The Weil algebra}

The above discussion leads to a simple and very useful example of a $\lieg$-differential algebra, named after Andr\'e Weil, which encodes abstractly the basic algebraic and differential properties of a connection and its curvature. If $S\lieg$ and $\Lambda\lieg$ denote the symmetric and exterior algebra over $\lieg$ respectively, the $\lieg$-Weil algebra is the tensor product
\be\label{Weilalg} \AW^{\lieg} = S\lieg\otimes\Lambda\lieg\, .\ee
It is generated by an odd element $\omega_{A}$ of ghost number one and an even element $\rho_{A}$ of ghost number two, both in the adjoint representation. The odd graded derivations $\dW$ and $i_{A}$ are defined by 
\begin{align}\label{dWeildef} &\dW\omega_{A} = \rho_{A} - \frac{1}{2}f_{ABC}\omega_{B}\omega_{C}\, ,\quad \dW\rho_{A} = f_{ABC}\rho_{B}\omega_{C}\\
\label{iWeildef}&i_{A}\omega_{B}=\delta_{AB}\, ,\quad i_{A}\rho_{B}=0\, .
\end{align}
It is straightforward to check that $(\AW^{\lieg},\dW,i_{A},l_{A})$  is a $\lieg$-differential algebra, with
\be\label{WeilLie} l_{A}\omega_{B}= f_{ABC}\omega_{C}\, ,\quad
l_{A}\rho_{B}= f_{ABC}\rho_{C}\, .\ee
The defining axioms \eqref{dWeildef} and \eqref{iWeildef} show that $\omega_{A}$ is a connection of curvature $\rho_{A}$. It is interesting to note that, if $(\A,D,\Int,\L)$ is any $\lieg$-differential algebra endowed with a connection $\theta$, then there always exists a unique morphism $\AW^{\lieg}\rightarrow\A$ mapping $\omega$ to $\theta$ and thus $\rho$ to the curvature \eqref{curvdef} of $\theta$, called the Chern-Weil morphism. Many results from the theory of characteristic classes, like the Weil transgression formula, etc., then have a direct analogue in the present abstract algebraic framework \cite{algebra}.

\subsubsection*{The standard BRST algebra}

Let $\mathscr A=\{\Phi\}$ be the algebra of physical local fields of a gauge theory of gauge group $G$, with gauge transformations
\be\label{gtgen} \delta\Phi = -\varepsilon_{A}L_{A}\Phi\, .\ee
The algebra
\be\label{BRSTAlg} \ABRST = \mathscr A\otimes\Lambda\lieg\ee
can be endowed with a $\lieg$-differential algebraic structure. The ghost number of the fields in $\mathscr A$ is set to zero and the generators $\Omega_{A}$ of $\Lambda\lieg$, called the ghosts, have $\gh\Omega_{A}=1$ and transform in the adjoint representation,
\be\label{ghosttrans}\L_{A}\Omega_{B}=f_{ABC}\Omega_{C}\, .\ee
The differential $s_{G}=s$ and interior products $I_{A}$ are defined by
\begin{align}\label{sdef} &s\Phi =  \Omega_{A}L_{A}\Phi\, ,\quad s\Omega_{A} = -\frac{1}{2}f_{ABC}\Omega_{B}\Omega_{C}\\\label{IBRSdef}
& I_{A}\Phi = 0\, ,\quad I_{A}\Omega_{B} = \delta_{AB}\, .
\end{align}
The axioms \eqref{ralg1}, \eqref{ralg2} are then trivially satisfied. Of course, the differential $s$ is the standard BRST operator. Note that the ghost $\Omega_{A}$ is a flat connection.

The minimal BRST algebra is often enlarged by including so-called ``trivial pairs,'' which are needed to build appropriate gauge-fixing fermions. A trivial pair $(\bar q^{I},\bar r^{I})$ transforms in some representation $(\tau_{A})^{I}_{\ J}$ of $G$,
\be\label{trivialp1} L_{A}\bar q^{I} = -(\tau_{A})^{I}_{\ J}\bar q^{J}\, ,\quad L_{A}\bar r^{I} = -(\tau_{A})^{I}_{\ J}\bar r^{J}\, .\ee
The elements $\bar q^{I}$ and $\bar r^{I}$ have opposite statistics, 
\be\label{epstrivial}\epsilon_{\bar q}=\epsilon_{\bar r}+1\ee
and ghost numbers such that
\be\label{ghostnumbertrivial} \gh\bar r = \gh\bar q +1\, .\ee
The definitions
\begin{align}\label{tp1} &s\bar q^{I} = -\bar r^{I}\, ,\quad s\bar r^{I} = 0
\\\label{tp2} &I_{A}\bar q^{I} = 0\, ,\quad I_{A}\bar r^{I} = (\tau_{A})^{I}_{\ J}\bar q^{J}
\end{align}
ensure that the axioms \eqref{ralg1} and \eqref{ralg2} are still satisfied. The fields $\bar q$ are usually called ``antighosts,'' whereas the $\bar r$'s are ``Lagrange multipliers.''

\subsection{Cohomology theories}\label{cohosec}

\subsubsection*{$\lieg$-differential algebras and cohomology theories}

Let $(\A,D,\Int_{A},\L_{A})$ a $\lieg$-differential algebra. We can naturally associate three cohomology rings to $\A$. The first ring $H^{*}(\A)$ is the cohomology ring of the differential $D$ acting on $\A$. The second ring $H^{*}(\A_{\text{inv}})$ is the cohomology ring of $D$ acting on the invariant subalgebra $\Ainv$, which is the subalgebra of $G$-invariant elements,
\be\label{Ainvdef} \Ainv = \bigcap_{A}\ker\L_{A}\, .\ee
This is well-defined, because $[\L_{A},D]=0$ and thus $\Ainv$ is stable under $D$. Finally, the last cohomology ring $H^{*}(\Abas)$ is associated with the so-called basic subalgebra
\be\label{Abasdef} \Abas =\bigcap_{A}\bigl(\ker\L_{A}\cap\ker\Int_{A}\bigr)\, ,\ee
which is also stable under $D$ thanks to the Cartan's identity. For example, in the case of the minimal BRST algebra defined by \eqref{BRSTAlg}--\eqref{IBRSdef}, $H^{*}(\ABRST)$ is the usual cohomology of the group $G$ with values in $\mathscr A$. If $\mathscr A=\mathbb R$ with the trivial $G$ action, we get the de Rham cohomology of $G$. On the other hand, $H^{*}(\A_{\text{BRST},\,\text{bas}})$ is simply the set of physical observables ($G$-invariant elements of $\mathscr A$).

\subsubsection*{The Weil model of equivariant cohomology}

The standard BRST algebra $\ABRST$ defined in \eqref{BRSTAlg}--\eqref{tp2}, with BRST operator $s_{G}=s$ computing the cohomology of the group $G$, is the basic framework in which the gauge-fixing procedure of a gauge field theory is usually discussed. If one wishes to fix the gauge only partially, from $G$ down to a non-trivial subgroup $H$, the r\^ole of the group $G$ is replaced by the quotient space $G/H$. Mathematically, going from $G$ to $G/H$ is equivalent to going from the standard cohomology of $G$ to the equivariant cohomology of $G$ with respect to $H$. It is thus extremely natural to guess that the $H$-equivariant version of the BRST framework  will be the correct tool to describe the partial gauge-fixing procedure. This will indeed be fully justified in the next section.

A simple algebraic description of equivariant cohomology is the so-called Weil model. One starts from a $\lieh$-differential algebra $(\A,s,I_{a},L_{a})$ and builds the tensor product algebra $(\A\otimes\AW^{\lieh},\sigma,\Int_{a},\L_{a})$ with the Weil algebra $(\AW^{\lieh},\dW,i_{a},l_{a})$ as explained around \eqref{dsum}--\eqref{Lsum}. The equivariant cohomology ring is then defined to be
\be\label{Weilequivring} H^{*}_{\text{Equiv}} = H^{*}\bigl((\A\otimes\AW^{\lieh})_{\text{bas}}\bigr)\, .\ee
For our purposes, we shall start with the usual BRST algebra associated with the cohomology of the group $G$. This is a $\lieg$-differential algebra and thus a $\lieh$-differential algebra as well. Decomposing the adjoint variables as in \eqref{Adjdec}, we find that the ghosts $\Omega_{a}$ and $\Omega_{i}$ transform in the adjoint and $R_{G/H}$ representations of $H$ respectively,
\be\label{ghHtrans} L_{a}\Omega_{b}=f_{abc}\Omega_{c}\, ,\quad L_{a}\Omega_{i}= f_{aij}\Omega_{j}\, .\ee
The relations \eqref{sdef} and \eqref{IBRSdef} read
\begin{align}\label{sW1} s\Phi &= \Omega_{a}L_{a}\Phi + \Omega_{i}L_{i}\Phi\\\label{rW2}
s\Omega_{a} &= -\frac{1}{2}f_{abc}\Omega_{b}\Omega_{c}-\frac{1}{2}f_{aij}\Omega_{i}\Omega_{j}\\\label{rW3}
s\Omega_{i} & = \Omega_{a}f_{aij}\Omega_{j}-\frac{1}{2}f_{ijk}\Omega_{j}\Omega_{k}\\\label{rW4}
I_{a}\Phi &= 0\, ,\quad I_{a}\Omega_{b}=\delta_{ab}\, ,\quad I_{a}\Omega_{i}=0\, .
\end{align}
Let us note that, even though the $\lieg$-connection $\Omega_{A}$ was flat, the $\lieh$-connection $\Omega_{a}$ has a non-zero curvature given by
\be\label{Rdef} R_{a} = -\frac{1}{2}f_{aij}\Omega_{i}\Omega_{j}\, .\ee
To build the Weil model, we add the generators $\omega_{a}$ and $\rho_{a}$ of the Weil algebra, which satisfy \eqref{dWeildef}--\eqref{WeilLie}. These new generators turn out to have very natural physical interpretations. For example, as will become clear in Sec.\ \ref{s3}, the $\rho_{a}$ play the r\^ole of ghosts for the ghosts $\Omega_{a}$. By definition, the 
operators $\sigma$, $\Int_{a}$ and $\L_{a}$ of the total space $\ABRST\otimes\AW^{\lieh}$ act on $\Phi$, $\Omega_{a}$ and $\Omega_{i}$ as in \eqref{ghHtrans}--\eqref{rW4} and on $\omega_{a}$ and $\rho_{a}$ as in \eqref{dWeildef}--\eqref{WeilLie}.

Since the relevant algebra for the equivariant cohomology \eqref{Weilequivring} is included in the kernel of the operators $\Int_{a}=I_{a}+i_{a}$, it is very convenient to make the change of variables
\be\label{changeofvar} \Omega_{a}\mapsto\Omega_{a}+\omega_{a}\, .\ee
The interior products $\Int_{a}$ then act trivially except on $\omega_{a}$,
\be\label{newvarI} \Int_{a}\omega_{b}=\delta_{ab}\, ,\quad \Int_{a}=0\ \text{on all the other generators}\, .\ee
Working in $\cap_{a}\ker\Int_{a}$ thus amounts to projecting out the  $\omega_{a}$. The basic subalgebra is then simply the $H$-invariant subspace of the so-called Cartan algebra
\be\label{Cartanalgdef} \AC = \ABRST\otimes S\lieg\, ,\ee
obtained from the usual BRST algebra by adding $\rho_{a}$.

For future reference, we list the action of $\sigma=s+\dW$ on the total algebra $\ABRST\otimes\AW^{\lieh}$, taking into account the change of variables \eqref{changeofvar},
\begin{align}\label{Dphi}& \sigma\Phi  = (\Omega_{a}+\omega_{a})\L_{a}\Phi
+ \Omega_{i}\L_{i}\Phi\\
\label{DOma} &\sigma\Omega_{a}  = R_{a} - \rho_{a}-\frac{1}{2}f_{abc}(\Omega_{b}\Omega_{c}+ 2 \Omega_{b}\omega_{c})\\
\label{DOmi} &\sigma\Omega_{i} = (\Omega_{a}+\omega_{a})f_{aij}\Omega_{j}-\frac{1}{2}f_{ijk}\Omega_{j}\Omega_{k}\\
\label{Doma}& \sigma\omega_{a}  = \rho_{a} - \frac{1}{2}f_{abc}\omega_{b}\omega_{c}\\
\label{Drhoa} &\sigma\rho_{a}  = \omega_{c}f_{cab}\rho_{b}\, .
\end{align}
The action of $\Int_{a}$ is given in \eqref{newvarI} and $\L_{a}=L_{a}+l_{a}$ acts in the natural way.

Let us mention that the transformation \eqref{changeofvar} is a special case of a more general transformation due to Kalkman \cite{kalkman} that can be defined for any algebra $\A\otimes\AW^{\lieh}$. There is actually an infinite family of Kalkman automorphisms $K_{t}$, parameterized by a real number $t$. In the BRST case, they act as $\Omega_{a}\mapsto \Omega_{a}+ t\omega_{a}$. In general, 
\be\label{kalkauto} K_{t} = e^{t\omega_{a}I_{a}}\, .\ee
It is straightforward to show that $K_{t}$ defined by this formula is an automorphism, in particular $K_{t}(xy)=K_{t}(x)K_{t}(y)$, and that the various operators transform as
\begin{align}\label{DKalk} K_{t}\sigma K_{t}^{-1} & = s + \dW + t(\omega_{a}L_{a}-\rho_{a}I_{a}) + \frac{1}{2}t(1-t)f_{abc}\omega_{b}\omega_{c}I_{a}\\
\label{IKalk} K_{t}\Int_{a}K_{t}^{-1} & = i_{a} + (1-t) I_{a}\\
\label{LKalk} K_{t}\L_{a}K_{t}^{-1} & = \L_{a} = l_{a} + L_{a}\, .
\end{align}
For $t=1$, one can easily check that \eqref{DKalk} is consistent with \eqref{Dphi}--\eqref{Drhoa} and that \eqref{IKalk} is consistent with \eqref{newvarI}.

\subsubsection*{The Cartan model of equivariant cohomology}

The point of view of Henri Cartan on equivariant cohomology is to work directly with the algebra \eqref{Cartanalgdef}. The advantage of this point of view is that the variable $\omega_{a}$, which is eventually projected out in the Weil model after the change of variables \eqref{changeofvar} is made,  is never introduced. The disadvantage is that $\AC$ is not an $\lieh$-differentiable algebra and in particular is not stable under the differential $\sigma$. One has to introduce on $\AC$ a new odd graded derivation of degree one, defined by
\be\label{dCdef} \dC = s-\rho_{a}I_{a}\, .\ee
This derivation is not a differential. Instead, it satisfies 
\be\label{dCsq} \dC^{2} = -\rho_{a}\L_{a}\, ,\quad [\dC,\L_{a}]=0\, .\ee
The square of $\dC$ thus vanishes on $H$-invariants; one says that $\dC$ is an \emph{equivariant differential}. Moreover, the invariant subalgebra of $\AC$ is stable under $\dC$. The associated cohomology ring is the Cartan model for the equivariant cohomology,
\be\label{Cartanequivring} H^{*}_{\text{Equiv}} = H^{*}\bigl(\A_{\text{C},\,\text{inv}}\bigr)\, .\ee
The equivalence with Weil's definition \eqref{Weilequivring} is very easy to establish. Indeed, the restriction of $\sigma$ to the Cartan subalgebra is given by
\be\label{DonCartan} \sigma_{\vert\AC} = \dC + \omega_{a}\L_{a}\, .\ee
This can be checked straightforwardly from \eqref{Dphi}--\eqref{Drhoa}, or equivalently from \eqref{DKalk} at $t=1$. Thus, $\sigma$ and $\dC$ coincide on $\AC$ up to gauge transformations belonging to $H$ and thus in particular are equal on the invariant subalgebra.

For future reference, we list the explicit formulas for the action of $\dC$,
\begin{align} \label{dCPhi}&\dC\Phi = \Omega_{a}\L_{a}\Phi + \Omega_{i}\L_{i}\Phi\\\label{dCOma} &\dC\Omega_{a} = R_{a}-\rho_{a}-\frac{1}{2}f_{abc}\Omega_{b}\Omega_{c}\\\label{dCOmi}&\dC\Omega_{i} = f_{aij}\Omega_{a}\Omega_{j}-\frac{1}{2}f_{ijk}\Omega_{j}\Omega_{k}\\\label{dCrhoa}
& \dC\rho_{a} = 0\\\label{dCRa} & \dC R_{a} = f_{abc}R_{b}\Omega_{c}\, .
\end{align}

\subsubsection*{Equivariant trivial pairs}

The minimal constructions discussed above are insufficient for our purposes, because all the variables have positive ghost numbers whereas the gauge-fixing fermion must have ghost number minus one. Of course, this is already true in the standard BRST framework, to which we must add trivial pairs $(\bar q,\bar r)$ satisfying \eqref{trivialp1}--\eqref{tp2}. Under the Kalkman automorphism $K_{1}$, see \eqref{kalkauto},
\be\label{Kalkmanonr} \bar r^{I}\mapsto \bar r^{I} + \omega_{a}(\tau_{a})^{I}_{\ J}\bar q^{J}\, .\ee
Using in particular \eqref{Doma} and \eqref{comab}, we can then compute
\begin{align} 
\label{Dqequiv}\sigma\bar q^{I} &= -\bar r^{I} -\omega_{a}(\tau_{a})^{I}_{\ J}\bar q^{J}=-\bar r^{I}+\omega_{a}\L_{a}\bar q^{I}\\\label{Drequiv} \sigma\bar r^{I} &= -(\tau_{a})^{I}_{\ J}\bigl(
 \rho_{a}\bar q^{J}+\omega_{a}\bar r^{J}\bigr) = -(\tau_{a})^{I}_{\ J}\rho_{a}\bar q^{J} + \omega_{a}\L_{a}\bar r^{I}\\\label{Iequiv} \Int_{a}\bar q^{I} & = 0\, ,\quad \Int_{a}\bar r^{I} = 0\, .
\end{align}
The action of the Cartan equivariant differential can also be found, e.g.\ from \eqref{DonCartan} and \eqref{Dqequiv}, \eqref{Drequiv},
\be\label{dCequiv} \dC\bar q^{I} = -\bar r^{I}\, ,\quad \dC\bar r^{I}
= -(\tau_{a})^{I}_{\ J}\rho_{a}\bar q^{J}\, .\ee
Pairs satisfying \eqref{epstrivial}, \eqref{ghostnumbertrivial} and \eqref{Dqequiv}--\eqref{Iequiv} (in the Weil model) or \eqref{dCequiv} (in the Cartan model) are called \emph{equivariant trivial pairs}.

In the following, we shall use two different equivariant trivial pairs: an $(\text{odd},\text{even})$ pair $(\bar\Omega_{i},\bar\La_{i})$ in the representation $R_{G/H}$ with ghost numbers $(-1,0)$; and an $(\text{even},\text{odd})$ pair $(\bar\rho_{a},\bar\La_{a})$ in the adjoint representation with ghost numbers $(-2,-1)$,
\begin{align}
\label{ex1tpD} & \sigma\bar\Omega_{i} = -\bar\La_{i}+f_{aij}\omega_{a}\bar\Omega_{j}\, ,\quad \sigma\bar\La_{i} = f_{aij}(\rho_{a}\bar\Omega_{j}+\omega_{a}\bar\La_{j})\\\label{ex2tpD}
& \sigma\bar\rho_{a} = -\bar\La_{a} + f_{abc} \omega_{c}\bar\rho_{b}\, ,\quad
\sigma\bar\La_{a} = f_{abc}(\rho_{c}\bar\rho_{b}+\omega_{c}\bar\La_{c})
\\\label{ex1tpdC} & \dC\bar\Omega_{i} = -\bar\La_{i}\, ,\quad
\dC\bar\La_{i} = f_{aij}\rho_{a}\bar\Omega_{j}\\\label{ex2tpdC} &
\dC\bar\rho_{a} = -\bar\La_{a}\, ,\quad\dC\bar\La_{a} = f_{abc}\rho_{c}\bar\rho_{b}\, .
\end{align}
\subsection{Summary}

On top of the physical fields $\Phi$, we use the ``ghosts'' $(\Omega_{i},\Omega_{a}, \omega_{a}, \rho_{a})$, ``antighosts'' $(\bar\Omega_{i},\bar\rho_{a})$ and ``Lagrange multipliers'' $(\bar\La_{i},\bar\La_{a})$. The pairs $(\bar\Omega_{i},\bar\La_{i})$ and $(\bar\rho_{a},\bar\La_{a})$ are equivariantly trivial. The parity and ghost numbers of all the variables are indicated in Table \ref{table}. The differential $\sigma$ acts as indicated in \eqref{Dphi}--\eqref{Drhoa} and \eqref{ex1tpD}, \eqref{ex2tpD}. Its square is zero and it commutes with the action of the gauge group $H$. The Cartan equivariant differential $\dC$ acts as in \eqref{dCPhi}--\eqref{dCrhoa} and \eqref{ex1tpdC}, \eqref{ex2tpdC} (it does not act on $\omega_{a}$). Its square \eqref{dCsq} is zero on $H$-invariant operators and it commutes with the action of $H$.

\begin{table}
\be\nonumber
\begin{matrix}
& \Phi & \Omega_{i} & \Omega_{a} &
\omega_{a} & \rho_{a} & \bar\Omega_{i} & \bar\La_{i} & \bar\rho_{a} & \bar\La_{a}\\
\hline
\text{parity }\epsilon & 0\ \text{or }1  & 1 & 1 & 1 & 0 & 1 & 0 & 0 & 1\\
\text{ghost number }\gh & 0  & 1 & 1 & 1 & 2 & -1 & 0 & -2 & -1
\end{matrix}\ee
\caption{\label{table} List of physical fields $\Phi$ (whose parity is 0 or 1 according to whether they are bosons or fermions), ghosts, antighosts and Lagrange multipliers used in the main text, with their parity and ghost numbers.}
\end{table}
\section{Equivariant gauge fixing}\label{s3}
\subsection{\label{sCartan} The Cartan model}

\subsubsection*{Cartan model and ghosts for ghosts}

From the physics point of view, the Cartan model can be interpreted as a ``ghosts for ghosts'' \cite{ghforgh} approach to the problem of partial gauge fixing. One starts with a ghost $\Omega_{A}$ suitable for the full gauge fixing of $G$. One then introduces ghosts for the ghosts $\Omega_{a}$, in order to actually gauge fix $G$ only down to $H$.
The Cartan variables $\rho_{a}$ have all the right properties to play the r\^ole of the ghosts for the ghosts, in particular their parity and ghost number match the expected values. We also need the antighosts $\bar\Omega_{i}$ and $\bar\rho_{a}$, which form equivariant trivial pairs $(\bar\Omega_{i},\bar\La_{i})$ and $(\bar\rho_{a},\bar\La_{a})$ with the Lagrange multipliers $\bar\La_{i}$ and $\bar\La_{a}$. The field content is exactly as in  Table \ref{table}, except for the Weil variable $\omega_{a}$, which plays no r\^ole in the Cartan construction.

\subsubsection*{The general Cartan ansatz for $S_{H}$}

Given a gauge-fixing fermion $\psi_{G/H}$ built from the above variables, of ghost number 
\be\label{psiGHgn} \gh\psi_{G/H}= -1\ee
and invariant under gauge transformations belonging to $H$,
\be\label{psiGHinv} \L_{a}\psi_{G/H} = 0\, ,\ee
we thus propose to define actions $S_{H}(\varphi;\psi_{G/H})$ by the path integral
\be\label{SHdefCartan} e^{-S_{H}(\varphi;\,\psi_{G/H})} = \int\!
D\phi D\Omega_{i}D\Omega_{a}D\rho_{a}D\bar\Omega_{i}D\bar\La_{i}
D\bar\rho_{a}D\bar\La_{a}\, e^{-S(\varphi,\phi) + \dC\psi_{G/H}}\, .
\ee
The actions $S_{H}$ so defined are manifestely invariant under $H$ gauge transformations. Moreover, we claim, and this will be justified below, that all these actions are physically equivalent. They all yield the same partition function as the original $G$-invariant gauge theory. However, let us emphasize that the actions $S_{H}$ themselves do depend on the choice of gauge-fixing fermions $\psi_{G/H}$. 

\subsubsection*{Building the gauge-fixing fermion}

In order to build an $H$-invariant fermion $\psi_{G/H}$, we need partial gauge-fixing conditions $F_{i}(\varphi,\phi)$ that transform covariantly under $H$. We thus impose that they belong to the representation $R_{G/H}$ defined by\eqref{realrep},
\be\label{condonFi} \L_{a}F_{i} = f_{aij}F_{j}\, .\ee
Finding partial gauge-fixing conditions satisfying \eqref{condonFi} is easy \cite{weinberg,tHooftAP}. For example, if $A^{\mu}_{A}$ is the gauge potential, one can check that the components $A^{\mu}_{a}$ and $A^{\mu}_{i}$ transform under $H$ as a gauge potential and in the representation $R_{G/H}$ respectively. One can then pick a $H$-covariant Lorenz-like gauge,
\be\label{Fiexample} F_{i} = \nabla_{\mu}A^{\mu}_{i} = \partial_{\mu}A^{\mu}_{i} + f_{aij}A_{\mu a}A^{\mu}_{j}\, ,\ee
using the covariant derivative $\nabla$ with respect to $H$.

The usual power-counting constraints from renormalization theory are also typically imposed. For example, in four dimensions, for standard gauge fixing functions $F_{i}$ of engineering dimension $[F_{i}]=2$ (as in \eqref{Fiexample}), the engineering dimensions of the various variables are given by
\be\label{engine} [\Omega_{i}]=[\Omega_{a}]=[\bar\Omega_{i}]=1\, ,\ [\bar\La_{i}]=[\rho_{a}]=[\bar\rho_{a}] = 2\, ,\ [\bar\La_{a}]=[\psi_{G/H}]=3\, .\ee
The most general gauge-fixing fermion then has the form
\be\label{psiGHform} \psi_{G/H} =\psi_{G/H}^{(1)} + \psi_{G/H}^{(2)} + \psi_{G/H}^{(3)}\, ,\ee
with
\begin{align}\label{psi1}& \psi_{G/H}^{(1)}= \bar\rho_{a}\Omega_{a}\\
\label{psi2} &\psi_{G/H}^{(2)} = \bar\Omega_{i}\Bigl(F_{i}(\varphi,\phi) - \frac{\xi}{2}\bar\La_{i}\Bigr)\\\label{psi3} &\psi_{G/H}^{(3)} = 
\frac{1}{2}\alpha f_{ijk}\bar\Omega_{i}\bar\Omega_{j}\Omega_{k} + \frac{1}{2}\tilde\alpha f_{ija}
\bar\Omega_{i}\bar\Omega_{j}\Omega_{a}\, .
\end{align}
It is convenient to assume in all cases that the full dependence on the equivariant trivial pair $(\bar\rho_{a},\bar\La_{a})$ comes from the term \eqref{psi1}. Computing
\be\label{dCpsi1} -\dC\psi_{G/H}^{(1)} = \bar\La_{a}\Omega_{a}+\bar\rho_{a}\Bigl(\rho_{a}-R_{a}+\frac{1}{2}f_{abc}\Omega_{b}\Omega_{c}\Bigr)\, ,\ee
we see that $\bar\La_{a}$ and $\bar\rho_{a}$ both play the r\^ole of Lagrange multipliers in the path integral \eqref{SHdefCartan}, imposing
\be\label{OmeRhoconstraints} \Omega_{a} = 0\, ,\quad \rho_{a} = R_{a} = -\frac{1}{2} f_{aij}\Omega_{i}\Omega_{j}\, .\ee
The ``ghosts for ghosts'' trick has allowed to eliminate $\Omega_{a}$ from the game. 

\subsubsection*{The restricted Cartan model}

It is then very natural to work with a restricted Cartan algebra $\ACh$ obtained from $\AC$ by imposing the constraints \eqref{OmeRhoconstraints}, which are compatible with the action of $\dC$. The algebra $\ACh$ is thus generated by the physical fields $\Phi$, ghosts $\Omega_{i}$ and equivariant trivial pair $(\bar\Omega_{i},\bar\La_{i})$. If 
\be\label{cansurject1} f:\AC\longrightarrow\ACh\ee
is the canonical surjective algebra homomorphism, we can define an equivariant differential $\dCh$ on $\ACh$ by
\be\label{dChabsdef} f\dC = \dCh f\, .\ee
Explicitly, we find
\begin{align}\label{dChonPhi} \dCh\Phi &= \Omega_{i}\L_{i}\Phi\\
\label{dChonOmi} \dCh\Omega_{i} &
= -\frac{1}{2}f_{ijk}\Omega_{i}\Omega_{j}\\
\label{dChonbarOmi} \dCh\bar\Omega_{i} &= -\bar\La_{i}\\
\label{dChonbarLai} \dCh\bar\La_{i} &= f_{aij}R_{a}\bar\Omega_{j}
=-\frac{1}{2}f_{aij}f_{akl}\Omega_{k}\Omega_{l}\bar\Omega_{j}\, .
\end{align}
The variable $R_{a}$ being defined by \eqref{OmeRhoconstraints}, we also find
\be\label{dChonRa} \dCh R_{a} = 0\ee
and the crucial identities
\be\label{dChrel} \dCh^{2} = -R_{a}\L_{a}\, ,\quad [\L_{a},\dCh]=0\, .\ee

\subsubsection*{The total gauge-fixed plus ghost action}

It is most efficient to work directly with the algebra $\ACh$. We thus integrate over $\bar\La_{a}$, $\bar\rho_{a}$, $\Omega_{a}$ and $\rho_{a}$ in \eqref{SHdefCartan} to get
\be\label{SHdef2} e^{-S_{H}(\varphi;\,\psi_{G/H})} = \int\! D\phi D\Omega_{i}
D\bar\Omega_{i}D\bar\La_{i}\, e^{-S(\varphi,\phi) + \dCh\psi_{G/H}(\varphi,\phi,\Omega_{i},\bar\Omega_{i},\bar\La_{i})}\, ,\ee
with a gauge-fixing fermion depending on two free parameters $\xi$ and $\alpha$,
\be\label{psigen} \psi_{G/H} = \bar\Omega_{i}\Bigl(F_{i}(\varphi,\phi) - \frac{\xi}{2}\bar\La_{i}\Bigr) + \frac{\alpha}{2} f_{ijk}\bar\Omega_{i}\bar\Omega_{j}\Omega_{k}\, .\ee
This is the most general four dimensional ansatz consistent with power counting for gauge-fixing conditions $F_{i}$ of engineering dimension two. Explicitly, the total partially gauge-fixed plus ghost action reads
\begin{multline}\label{dChonpsi}S_{\text{tot}} = S-\dCh\psi_{G/H} = S(\varphi,\phi)-\frac{\xi}{2}\bar\La_{i}\bar\La_{i}
+\bar\La_{i}\bigl(F_{i}(\varphi,\phi) + \alpha f_{ijk}\bar\Omega_{j}\Omega_{k}\bigr)
+ \bar\Omega_{i}\L_{j}F_{i}\Omega_{j} \\+ 
\frac{1}{4}\bigl(\xi f_{aij}f_{akl}+\alpha f_{mij}f_{mkl}\bigr)
\Omega_{i}\Omega_{j}\bar\Omega_{k}\bar\Omega_{l}\, .
\end{multline}
We can go on and integrate out $\bar\La_{i}$, producing the Gaussian weight $\frac{1}{2\xi}F_{i}F_{i}$ together with an additional quartic ghost term and a $\bar\Omega\Omega F$ coupling,
\begin{multline}\label{Stotfinal} S_{\text{tot}} = S(\varphi,\phi) + \frac{F_{i}F_{i}}{2\xi} + \bar\Omega_{i}\Bigl(\L_{j}F_{i} + \frac{\alpha}{\xi} f_{ijk}F_{k}\Bigr)\Omega_{j}\\ + \frac{1}{4}\Bigl(
\xi f_{aij}f_{akl} + \alpha f_{mij}f_{mkl} + \frac{2\alpha^{2}}{\xi} f_{mjk}f_{mil}\Bigr) \Omega_{i}\Omega_{j}\bar\Omega_{k}\bar\Omega_{l}\, .
\end{multline}
As far as practical calculations are concerned, the formulas \eqref{dChonpsi} and \eqref{Stotfinal} are the most important of our work. One can directly start from them to compute the effective action $S_{H}$.

It is interesting to note that the formulas simplify considerably when $G/H$ is a symmetric space. In this case $f_{ijk}=0$ and thus
\be\label{Stotsymspace} S_{\text{tot}} = S(\varphi,\phi) + 
\frac{F_{i}F_{i}}{2\xi} + \bar\Omega_{i}\L_{j}F_{i}\Omega_{j}+
\frac{\xi}{4}f_{aij}f_{akl}\Omega_{i}\Omega_{j}\bar\Omega_{k}\bar\Omega_{l}\, .\ee

The most salient qualitative feature of the results \eqref{dChonpsi}--\eqref{Stotsymspace} is to show that quartic ghost coupling must be included. \emph{This is an unavoidable property of the partial gauge-fixing procedure and it is a purely tree-level effect, independent of any consideration of renormalizability.} The only exception would be the case $\xi=\alpha=0$ (or simply $\xi=0$ when $G/H$ is a symmetric space), but this corresponds to a non-renormalizable gauge \`a la Landau where the gauge-fixing conditions $F_{i}=0$ are imposed strictly via a $\delta$-function in the path integral.

Let us now turn to the proof of the crucial claim that the partition functions computed from the actions $S_{H}$ defined by \eqref{SHdefCartan} of equivalently \eqref{SHdef2} all match the partition function of the original $G$-invariant gauge theory. This actually follows quite simply from the Weil point of view on equivariant cohomology.

\subsection{\label{sWeil} The Weil model and gauge-fixing independence}

\subsubsection*{The restricted Weil model}

It is convenient to introduce a restricted Weil model for which the constraints \eqref{OmeRhoconstraints}, $\Omega_{a}=0$ and $\rho_{a}=R_{a}$, are imposed. These constraints are compatible with the action \eqref{Dphi}--\eqref{Drhoa} and \eqref{Dqequiv}, \eqref{Drequiv} of $\sigma$. The corresponding  restricted Weil algebra, generated by the physical fields $\Phi$, ghosts $\Omega_{i}$ and $\omega_{a}$ and equivariant trivial pair $(\bar\Omega_{i},\bar\La_{i})$ is thus endowed with a differential $\hat\sigma$ descending from $\sigma$. Explicitly, if $\mathsf f$ is the canonical surjective morphism from the total algebra $\ABRST\otimes\AW^{\lieh}$ of the Weil model onto the restricted algebra, $\hat\sigma$ is defined by (compare with \eqref{dChabsdef})
\be\label{SigmaDelta} \mathsf f \sigma = \hat\sigma \mathsf f\, .\ee
Explicitly, from \eqref{Dphi}--\eqref{Doma} and \eqref{ex1tpD}, \eqref{ex2tpD}, we find
\begin{align}
\label{DelonPhi} &\hat\sigma\Phi  = \Omega_{i}\L_{i}\Phi+ \omega_{a}\L_{a}\Phi
\\\label{DelonOmei}& \hat\sigma\Omega_{i}  = -\frac{1}{2}f_{ijk}\Omega_{j}\Omega_{k} + f_{aij}\omega_{a}\Omega_{j}\\\label{Delonoma} &
\hat\sigma\omega_{a}  = R_{a} - \frac{1}{2}f_{abc}\omega_{b}\omega_{c}\\ 
\label{DelonbarOmi} & \hat\sigma\bar\Omega_{i} = -\bar\La_{i} + f_{aij}
\omega_{a}\bar\Omega_{j}\\\label{DelonbarLai} & \hat\sigma\bar\La_{i}  =
f_{aij}\bigl(R_{a}\bar\Omega_{j} + \omega_{a}\bar\La_{j}\bigr)\, .
\end{align}
We also have
\be\label{DelonRa} \hat\sigma R_{a} = f_{abc}\omega_{c}R_{b}\, .\ee
Moreover, on the restricted Cartan algebra $\ACh$,
\be\label{Deldelrel} \hat\sigma_{\vert\ACh} = \dCh + \omega_{a}\L_{a}\, ,\ee
which is the analogue of the relation \eqref{DonCartan}.

It is useful to extend the action of $\hat\sigma$ to a new ordinary trivial pair $(\bar\omega_{a},\bar\la_{a})$ in the adjoint of $H$,
\be\label{Delontpa} \hat\sigma\bar\omega_{a} = -\bar\la_{a}\, ,\quad
\hat\sigma\bar\la_{a} = 0\, .\ee
If we denote by $\AWeil$ the algebra generated by all the above variables, it is easy to check that $(\AWeil,\hat\sigma,\Int_{a},\L_{a})$ is an $\lieh$-differential algebra, the interior products $\Int_{a}$ acting trivially except for
\be\label{IresWeildef} \Int_{a}\omega_{b} = \delta_{ab}\, ,\quad
\Int_{a}\bar\La_{b} = -f_{abc}\bar\omega_{c}\, .\ee

Most interestingly, the restricted Weil algebra $(\AWeil,\hat\sigma,\Int_{a},\L_{a})$ constructed above is actually isomorphic to the standard BRST algebra for the group $G$, viewed as an $\lieh$-differential algebra, with ghosts $\Omega_{A}=(\omega_{a},\Omega_{i})$, $\hat\sigma$ and $\Int_{a}$ being identified with the standard BRST operator $s$ and interior product $I_{a}$ acting as in \eqref{sdef}, \eqref{IBRSdef}, \eqref{tp1} and \eqref{tp2}. Indeed, the action \eqref{DelonPhi}--\eqref{Delonoma} and \eqref{Delontpa} of $\hat\sigma$ trivially matches the action of the standard BRST operator on $\Phi$, $\Omega_{i}$, $\omega_{a}$, $\bar\omega_{a}$ and $\bar\la_{a}$. The only apparent mismatch is for the action \eqref{DelonbarOmi}, \eqref{DelonbarLai} on $(\bar\Omega_{i},\bar\La_{i})$, which behaves as an equivariant trivial pair and not as an ordinary BRST trivial pair. However, we know from our discussion of the Kalkman automorphism in Sec.\ \ref{cohosec} that this can be undone by a suitable change of variables \eqref{Kalkmanonr}. If we note
\be\label{newbarlai}\bar\la_{i} = \bar\La_{i} - f_{aij}\omega_{a}\bar\Omega_{j}\, ,\ee
it is indeed straightforward to check that $(\bar\Omega_{i},\bar\la_{i})$ is an ordinary trivial pair for $\hat\sigma$,
\begin{align}\label{Delonlittlela}& \hat\sigma\bar\Omega_{i} = -\bar\la_{i}\, ,\quad\hat\sigma\bar\la_{i} = 0\\\label{Iintonlittlela}&
\Int_{a}\bar\Omega_{i} = 0\, , \quad \Int_{a}\bar\la_{i} = -f_{aij}\bar\Omega_{j}\, .
\end{align}
These simple observations will be extremely useful in the next subsubsection. 

\subsubsection*{The fundamental theorem}

We now have all the tools to give a simple proof of the following fundamental theorem.

\noindent\textsc{Theorem}: \emph{Let $S_{H}(\varphi;\psi_{G/H})$ be the $H$-invariant action defined by \eqref{SHdef2} and let}
\be\label{Zpsipsidef} \mathscr Z(\psi_{H},\psi_{G/H}) = \int\! D\varphi D\omega_{a}D\bar\omega_{a} D\bar\la_{a}\, e^{-S_{H}(\varphi;\,\psi_{G/H}) + s_{H}\psi_{H}(\varphi,\omega_{a}\bar\omega_{a},\bar\la_{a})}\, ,\ee
\emph{where $s_{H}$ is the usual BRST operator for the gauge group $H$, acting on the fields $\varphi$, ghosts $\omega_{a}$ and trivial pair $(\bar\omega_{a},\bar\la_{a})$. Then
\be\label{ZscriptZ} \mathscr Z(\psi_{H},\psi_{G/H})=Z\ee
is the partition function of the gauge theory defined by \eqref{Zdef}. In particular, the variations of $\mathscr Z$ with respect to the gauge-fixing fermions $\psi_{H}$ and $\psi_{G/H}$ vanish.}

This theorem ensures that all the $H$-invariant actions $S_{H}(\varphi,\psi_{G/H})$, parameterized by $\psi_{G/H}$, really define the same gauge theory, which is the original $G$-invariant model; after standard gauge fixing of the $H$ gauge symmetry, all the actions $S_{H}$ reproduce the same partition function $Z$. This fully justifies the general formulas \eqref{SHdefCartan} or \eqref{SHdef2}, which were proposed based on the abstract Cartan construction of equivariant cohomology in the previous subsection. We can prove the theorem in three simple steps.

\noindent\emph{Step one}: if $\psi_{H}$ and $\psi'_{H}$ are sufficiently close to each other,
\be\label{step1} \mathscr Z(\psi_{H},\psi_{G/H}) = 
\mathscr Z(\psi_{H}',\psi_{G/H})\, .\ee
This is the usual gauge-fixing independence of the standard BRST framework. It is ensured by the $H$-invariance of $S_{H}$, in particular by
\be\label{sHonSH} s_{H}S_{H}(\varphi) = \omega_{a}\L_{a}S_{H}(\varphi) = 0\, .\ee

\noindent\emph{Step two}: we use the step one to pick
\be\label{psiprime}\psi_{H}' = \bar\omega_{a}F_{a}(\varphi)\, ,\ee
where $F_{a}$ is a set of standard gauge-fixing conditions for $H$. In particular, the Hessian
\be\label{Heedef} H_{ab}(\varphi) = \L_{b}F_{a}\ee
in an invertible matrix. Since
\be\label{sonpsiprime} -s_{H}\psi_{H}' = \bar\la_{a}F_{a}+\bar\omega_{a}\L_{b}F_{a}\omega_{b} \, ,\ee
we have
\begin{multline}\label{Zpr1}\mathscr Z (\psi_{H},\psi_{G/H}) = 
\int\! D\varphi D\omega_{a}D\bar\omega_{a} D\bar\la_{a}
D\phi D\Omega_{i} D\bar\Omega_{i}D\bar\La_{i}\\
e^{-S(\varphi,\phi) -\bar\la_{a}F_{a}-\bar\omega_{a}\L_{b}F_{a}\omega_{b}  +\dCh\psi_{G/H}(\varphi,\phi,\Omega_{i},\bar\Omega_{i}\bar\La_{i})}\, .
\end{multline}
Let us then make the change of variables 
\be\label{changeomega} \omega_{a} \mapsto \omega_{a}+H^{-1}_{ab}(\varphi)\L_{i}F_{b}\,\Omega_{i}\ee
in the path integral \eqref{Zpr1}. Noting that $\dCh\psi_{G/H}$ does not depend on $\omega_{a}$ and that
\be\label{Delonpsiprime} -\hat\sigma\psi'_{H} = \bar\la_{a}F_{a}+\bar\omega_{a}\L_{b}F_{a}\omega_{b}+\bar\omega_{a}\L_{i}F_{a}\Omega_{i}\, ,\ee
we find
\be\label{Zpr2}\mathscr Z (\psi_{H},\psi_{G/H}) = 
\int\! D\varphi D\omega_{a}D\bar\omega_{a} D\bar\la_{a}
D\phi D\Omega_{i} D\bar\Omega_{i}D\bar\La_{i}\,
e^{-S(\varphi,\phi) +\hat\sigma\psi'_{H} + \dCh\psi_{G/H}}\, .\ee

\noindent\emph{Step three}: we now use \eqref{Deldelrel}, the $H$-invariance of the fermion $\psi_{G/H}$ and the fact that it depends only on variables in the restricted Cartan algebra to get
\be\label{dDelonPsi} \dCh\psi_{G/H} = \hat\sigma\psi_{G/H}\, .\ee
The non-trivial point in this identity is that all the $\omega_{a}$-dependent terms in the right-hand side of the equation cancel each other. Thus we have
\be\label{Zpr3}\mathscr Z (\psi_{H},\psi_{G/H}) = 
\int\! D\varphi D\omega_{a}D\bar\omega_{a} D\bar\la_{a}
D\phi D\Omega_{i} D\bar\Omega_{i}D\bar\La_{i}\,
e^{-S(\varphi,\phi) +\hat\sigma (\psi'_{H} + \psi_{G/H})}\, .\ee
Since, as explained at the end of the previous subsubsection, $\hat\sigma$ is acting as the standard BRST operator for the gauge group $G$, up to the change of variable \eqref{newbarlai}, we see that the equation \eqref{Zpr3} is exactly of the form \eqref{Zdef}. This concludes the proof.\hfill$\Box$

%
\section{Elementary approach}\label{s4}

We are now going to start our analysis anew, independently of the discussion of the previous section. Our aim is to explain all the results in a completely elementary way. In a sense, our presentation can be seen as a dissection of the abstract formalism that we have used up to now.

\subsection{Diagonalizing the ghost Lagrangian}\label{diagosec}

\subsubsection*{General discussion}

Let us start directly from the definition of the partition function \eqref{Zdef}. A typical choice for the gauge-fixing fermion is
\be\label{gffstandard} \Psi_{G} = \bar\omega_{A}\Bigl(F_{A}(\Phi) -\frac{\xi}{2}\bar\la_{A}\Bigr)\, ,\ee
where $F_{A}$, $1\leq A\leq\dim G$, are the gauge fixing conditions. Since we aim at fixing the gauge only partially down to $H$, we decompose the index $A=(a,i)$ according to \eqref{Liedec} or \eqref{Adjdec} and use gauge-fixing conditions $F_{i}$ that transform covariantly under $H$,
\be\label{condonFibis} \L_{a}F_{i} = f_{aij}F_{j}\, .\ee
This is as in \eqref{condonFi}. We always note $\L_{A}$ the action of the generator $\tau_{A}$ of the gauge group. Moreover, we use gauge-fixing conditions $F_{a}$ for $H$ that depend only on the fields $\varphi$ in the decomposition \eqref{phiPhisep}. It is also natural in this context to slightly generalize the $\xi\bar\omega_{A}\bar\la_{A}$ term in \eqref{gffstandard} to $\zeta\bar\omega_{a}\bar\la_{a}+\xi\bar\omega_{i}\bar\la_{i}$, preserving the $H$-invariance. We thus consider
\be\label{gfGimpro} \Psi_{G} = \bar\omega_{a}\Bigl(F_{a}(\varphi) - \frac{\zeta}{2}\bar\la_{a}\Bigr) + \bar\omega_{i}\Bigl(F_{i}(\Phi) - \frac{\xi}{2}\bar\la_{i}\Bigr)\, .\ee
Written in this way, one might believe that the $H$-invariant second term in $\Psi_{G}$ implements the partial gauge-fixing $G\rightarrow H$ that we seek, the first term being added on a second stage to gauge fix $H\rightarrow\{\text{Id}\}$. However, this is misleading. Acting with the BRST operator $s_{G}$ indeed yields (see Sec.\ \ref{gdiffsec} for a review of the standard BRST algebra)
\begin{multline}\label{sGonpsistan} -s_{G}\Psi_{G} = \bar\la_{a}F_{a}- \frac{\zeta}{2}\bar\la_{a}\bar\la_{a} + \bar\omega_{a}\L_{b}F_{a}\omega_{b}
 + \bar\la_{i}F_{i} - \frac{\xi}{2}\bar\la_{i}\bar\la_{i}
+\bar\omega_{i}\L_{j}F_{i}\omega_{j}\\
+ \bar\omega_{a}\L_{i}F_{a}\omega_{i} + \bar\omega_{i}\L_{a}F_{i}\omega_{a}\, .
\end{multline}
The ``off-diagonal'' terms in the second line of the above equation mix up variables which are na\"\i vely associated with the two steps $G\rightarrow H$ and $H\rightarrow \{\text{Id}\}$ in the gauge fixing and prevent the decomposition of the path integral as in \eqref{ZHdef}, \eqref{SHdef}. What is needed is a trick to diagonalize the ghost Lagrangian in the $(a,i)$ indices.

\subsubsection*{A simple special case: $\xi=0$} 

When $\xi=0$, which is the case that was treated correctly in \cite{weinberg}, the diagonalization is actually trivial to perform. One first uses the change of variables
\be\label{bigLadef} \bar\La_{i} = \bar\la_{i} + f_{aij}\omega_{a}\bar\omega_{j}\ee
which, by using \eqref{condonFibis}, eliminates the $\bar\omega_{i}\omega_{a}$ term from \eqref{sGonpsistan}. One then redefines
\be\label{omeganewdef} \omega'_{a} = \omega_{a} + H^{-1}_{ab}\L_{i}F_{b}\omega_{i}\, ,\ee
where $H^{-1}$ is the inverse matrix of the Hessian $H_{ab}=\L_{b}F_{a}$ (this Hessian is always invertible for good gauge-fixing conditions $F_{a}$). This eliminates the $\bar\omega_{a}\omega_{i}$ term and thus achieves the diagonalization,
\be\label{sGonp2} -s_{G}\Psi_{G} = \bar\la_{a}F_{a}- \frac{\zeta}{2}\bar\la_{a}\bar\la_{a} + \bar\omega_{a}\L_{b}F_{a}\omega'_{b}
 + \bar\La_{i}F_{i} 
+\bar\omega_{i}\L_{j}F_{i}\omega_{j}\, .\ee
If we then define 
\be\label{SHtriv}\begin{split} e^{-S_{H}(\varphi)}&= \int\! D\phi D\omega_{i}D\bar\omega_{i}D\bar\La_{i}\, e^{-S(\varphi,\phi) - \bar\La_{i}F_{i} - \bar\omega_{i}\L_{j}F_{i}\omega_{j}}
\\ & = \int\! D\phi D\omega_{i}D\bar\omega_{i}\,\delta(F_{i})\, e^{-S(\varphi,\phi) - \bar\omega_{i}\L_{j}F_{i}\omega_{j}}\, ,
\end{split}\ee
the partition function is given by the formula \eqref{ZHdef}, $\omega_{a}$ being renamed as $\omega'_{a}$, with
\be\label{psiHex} \psi_{H} = \bar\omega_{a}\Bigl(F_{a}(\varphi) - \frac{\zeta}{2}\bar\la_{a}\Bigr)\, .\ee

We have thus obtained one possible solution to our problem. Its main drawback is that it corresponds to a non-renormalizable gauge, the gauge-fixing conditions $F_{i}=0$ being imposed strictly in the path integral via a $\delta$-function. This is usually unacceptable for practical calculations.

\subsubsection*{The general case}

When $\xi\not = 0$, we can still use the change of variable \eqref{bigLadef} to eliminate the $\bar\omega_{i}\omega_{a}$ term in \eqref{sGonpsistan}, but the $-\frac{\xi}{2}\bar\la_{i}\bar\la_{i}$ term then yields even more complicated-looking mixed cubic and quartic terms,
\begin{multline}\label{sGpsiGcom} -s_{G}\Psi_{G} = \bar\la_{a}F_{a}-\frac{\zeta}{2}\bar\la_{a}\bar\la_{a} + \bar\omega_{a}\L_{b}F_{a}\omega_{b} +
\bar\omega_{a}\L_{i}F_{a}\omega_{i} + \bar\La_{i}F_{i}-\frac{\xi}{2}\bar\La_{i}\bar\La_{i} + \bar\omega_{i}\L_{j}F_{i}\omega_{j}\\
+ \xi\Bigl( f_{aij}\bar\La_{i}\omega_{a}\bar\omega_{j} + \frac{1}{2}
f_{aik}f_{bjk}\omega_{a}\omega_{b}\bar\omega_{i}\bar\omega_{j}\Bigr)\, .
\end{multline}
This is where a miracle occurs. The unwanted mixed terms on the second line of \eqref{sGpsiGcom} can be canceled by adding the cubic ghost term
\be\label{addtoPsiG} \frac{\xi}{2}f_{aij}\omega_{a}\bar\omega_{i}\bar\omega_{j}\ee
in the gauge-fixing fermion $\Psi_{G}$, \emph{at the expense of producing new crucial quartic ghost couplings involving only $\omega_{i}$ and $\bar\omega_{i}$}. Of course, from the standard BRST argument, adding \eqref{addtoPsiG} to $\Psi_{G}$ does not change the partition function $Z$. By carefully using the Jacobi identity and the relation \eqref{bigLadef}, we find precisely
\begin{multline}\label{sGonpsi3} -s_{G}\Bigl(\Psi_{G} + \frac{\xi}{2}f_{aij}\omega_{a}\bar\omega_{i}\bar\omega_{j}\Bigr) =
\bar\la_{a}F_{a}-\frac{\zeta}{2}\bar\la_{a}\bar\la_{a} + \bar\omega_{a}\L_{b}F_{a}\omega_{b} +
\bar\omega_{a}\L_{i}F_{a}\omega_{i}\\ + \bar\La_{i}F_{i}-\frac{\xi}{2}\bar\La_{i}\bar\La_{i} + \bar\omega_{i}\L_{j}F_{i}\omega_{j}
+ \frac{\xi}{4}f_{aij}f_{akl}\omega_{i}\omega_{j}\bar\omega_{k}\bar\omega_{l}\, .
\end{multline}
We can then use the redefinition \eqref{omeganewdef} to complete the diagonalization of the ghost Lagrangian,
\begin{multline}\label{sGonpsi4} -s_{G}\Bigl(\Psi_{G} + \frac{\xi}{2}f_{aij}\omega_{a}\bar\omega_{i}\bar\omega_{j}\Bigr) =
\bar\la_{a}F_{a}-\frac{\zeta}{2}\bar\la_{a}\bar\la_{a} + \bar\omega_{a}\L_{b}F_{a}\omega'_{b} \\ + \bar\La_{i}F_{i}-\frac{\xi}{2}\bar\La_{i}\bar\La_{i} + \bar\omega_{i}\L_{j}F_{i}\omega_{j}
+ \frac{\xi}{4}f_{aij}f_{akl}\omega_{i}\omega_{j}\bar\omega_{k}\bar\omega_{l}\, .
\end{multline}
Actually, this is not the most general form for the diagonal gauge-fixed action. We can add an additional term
\be\label{newnewnew} \frac{\alpha}{2}f_{ijk}\bar\omega_{i}\bar\omega_{j}\omega_{k}\ee
to $\Psi_{G}$ and show, using again \eqref{bigLadef} and the Jacobi identity, that all the terms containing $\omega_{a}$ cancel in the expression obtained by acting with $s_{G}$. The resulting precise dependence in $\alpha$ is indicated in Eq.\ \eqref{Stotelem} below.

\subsubsection*{Summary}

We can summarize our findings as follows. By choosing the gauge-fixing fermion $\Psi_{G}$ in the path integral \eqref{Zdef} defining the partition function as
\be\label{psiGgen} \Psi_{G} = \bar\omega_{a}\Bigl(F_{a}(\varphi) - \frac{\zeta}{2}\bar\la_{a}\Bigr) + \bar\omega_{i}\Bigl(F_{i}(\Phi) - \frac{\xi}{2}\bar\la_{i}\Bigr) + \frac{1}{2}\bigl(\xi f_{aij}\omega_{a}\bar\omega_{i}\bar\omega_{j} + \alpha f_{ijk}\bar\omega_{i}\bar\omega_{j}\omega_{k}\bigr)\ee
and using the change of variables \eqref{bigLadef} and \eqref{omeganewdef}, we find
\be\label{sGaction} S(\Phi) - s_{G}\Psi_{G} = -s_{H}\psi_{H} + S_{\text{tot}}\, ,\ee
where $s_{H}$ is the BRST differential for the group $H$ acting in the usual way on $(\Phi,\omega'_{a},\bar\omega_{a},\bar\la_{a})$,
\be\label{psiH} \psi_{H} = \bar\omega_{a}\Bigl(F_{a}(\varphi) - \frac{\zeta}{2}\bar\la_{a}\Bigr)\, ,\ee
and
\begin{multline}\label{Stotelem} S_{\text{tot}} = S(\varphi,\phi) 
-\frac{\xi}{2}\bar\La_{i}\bar\La_{i}
+\bar\La_{i}\bigl(F_{i}(\varphi,\phi) + \alpha f_{ijk}\bar\omega_{j}\omega_{k}\bigr)
+ \bar\omega_{i}\L_{j}F_{i}\omega_{j} \\+ 
\frac{1}{4}\bigl(\xi f_{aij}f_{akl}+\alpha f_{mij}f_{mkl}\bigr)
\omega_{i}\omega_{j}\bar\omega_{k}\bar\omega_{l}\, .
\end{multline}
If we then define the manifestly $H$-invariant action $S_{H}$ by
\be\label{SHelemdef} e^{-S_{H}(\varphi)} = \int\! D\phi D\omega_{i}
D\bar\omega_{i}D\bar\La_{i}\, e^{-S_{\text{tot}}}\ee
then, by construction, the partition function is automatically given by \eqref{ZHdef} (renaming $\omega'_{a}$ the dummy variable $\omega_{a}$). This yields a fairly general solution to our problem, that allows in particular to deal with a Gaussian weight for the partial gauge-fixing conditions. Let us emphasize that the action $S_{H}$ defined by \eqref{SHelemdef} does depend on the parameters $\xi$, $\alpha$ and on the choice of the conditions $F_{i}$. However, and this is the crucial point, all these different possible actions $S_{H}$ are strictly equivalent, since they yield by construction the same full partition function $Z$.

Some of the most important results of Section \ref{s3} have been reproduced. The formula \eqref{Stotelem} for the total action matches the formula \eqref{dChonpsi} obtained from the Cartan model of equivariant cohomology (the ghosts and antighosts were denoted by $\Omega_{i}$ and $\bar\Omega_{i}$ in Sec.\ \ref{s3} instead of $\omega_{i}$ and $\bar\omega_{i}$ presently). The miraculous cancellation of all the mixed, $\omega_{a}$-dependent terms achieved by adding the contribution \eqref{addtoPsiG} to $\Psi_{G}$ is automatically implemented in the algebraic framework. The new terms are generated by the equivariant trivial pairs transformation rules, e.g.\ \eqref{DelonbarOmi}, \eqref{DelonbarLai}, and the $\omega_{a}$-independence is ensured by identities like \eqref{Deldelrel}. The $\omega_{a}$-independence of the terms generated by acting with the BRST differential on \eqref{newnewnew} is similarly implied by \eqref{Deldelrel}. These properties will be further clarified below.

\subsection{The equivariant differential}

The partial gauge-fixing procedure just described turns out to be governed by an analogue of the usual BRST structure. This is not obvious, at first sight, because the usual BRST differential $s_{G}$ does not act in a closed form on the fields $\Phi$ or on the ghosts $\omega_{i}$,
\begin{align} \label{sonPhi} s_{G}\Phi &= \omega_{i}\L_{i}\Phi + \omega_{a}\L_{a}\Phi\\\label{sonOmei} s_{G}\omega_{i} & = -\frac{1}{2}f_{ijk}\omega_{j}\omega_{k} -f_{ija}\omega_{j}\omega_{a} = -\frac{1}{2}f_{ijk}\omega_{j}\omega_{k} + \omega_{a}\L_{a}\omega_{i}\, .
\end{align}
Of course, the problem can be traced to the fact that the ``algebra'' generated by the $\tau_{i}$ that we want to gauge fix is not closed. In this sense, our problem bears some similarity with the usual problem of gauge fixing open gauge algebras (for a nice review on this subject, see e.g.\ \cite{henneaux}). In our case, the ``trivial'' gauge transformations are the $H$-gauge transformations. The difference from the usual open gauge algebras is that $H$ is not a normal subgroup of $G$, except in the trivial instance where we have a product group structure. Moreover, in our case the actions $S_{H}$ we want to build are not independent of the gauge-fixing fermion that implements the partial gauge fixing from $G$ down to $H$. Only the full partition function $Z$ defined by \eqref{ZHdef} should be.

This being said, the transformation rules \eqref{sonPhi} and \eqref{sonOmei} are quite suggestive. Since the modified BRST transformations we seek will act on a gauge-fixing fermion which must be invariant under the gauge transformations belonging to $H$, we can a priori forget about the terms in $\omega_{a}\L_{a}$ in \eqref{sonPhi} and \eqref{sonOmei} and tentatively define an odd graded derivation $\delta$ by
\be\label{deltaact1} \delta\Phi = \omega_{i}\L_{i}\Phi\, ,\quad
\delta\omega_{i} = -\frac{1}{2}f_{ijk}\omega_{j}\omega_{k}\, .\ee
It is then straightforward to check that the square of $\delta$ does not vanish in general but instead satisfies
\be\label{deltasq} \delta^{2} = -R_{a}\L_{a}\, ,\ee
where the so-called ``curvature'' $R_{a}$ is defined by (see also \eqref{Rdef})
\be\label{Radefnew} R_{a} = -\frac{1}{2}f_{aij}\omega_{i}\omega_{j}\, .\ee
One also has the useful identities
\be\label{deltaonR} \delta R_{a} = 0\ee
and
\be\label{deltaLa} [\L_{a},\delta] = 0\, .\ee
The relations \eqref{deltasq} and \eqref{deltaLa} show that $\delta$ is an equivariant differential with respect to $H$. In particular, its square vanishes on $H$-invariants. 

Let us now introduce the trivial pair $(\bar\omega_{i},\bar\la_{i})$ on which the BRST differential acts as $s_{G}\bar\omega_{i} = -\bar\la_{i}$ and $s_{G}\bar\la_{i} = 0$. Using the Jacobi identity, one finds from the change of variables \eqref{bigLadef} that
\begin{align}\label{sGtbarom} s_{G}\bar\omega_{i} &= -\bar\La_{i} 
+ f_{aij}\omega_{a}\bar\omega_{j} = -\bar\La_{i} + \omega_{a}\L_{a}\bar\omega_{i}\\
s_{G}\bar\La_{i} &= -\frac{1}{2}f_{aij}f_{akl}\omega_{k}\omega_{l}\bar\omega_{j} + f_{aij}\omega_{a}\bar\La_{j} = f_{aij}R_{a}\bar\omega_{j} + \omega_{a}\L_{a}\bar\La_{i}\, .
\end{align}
Following the same idea that led to the definitions \eqref{deltaact1} from \eqref{sonPhi} and \eqref{sonOmei}, we set
\be\label{deltaact2} \delta\bar\omega_{i} = -\bar\La_{i}\, ,\quad
\delta\bar\La_{i} = f_{aij}R_{a}\bar\omega_{j}\, .\ee
One can then check that these definitions are consistent with the fundamental identities \eqref{deltasq} and \eqref{deltaLa}. 

The equivariant differential $\delta$ that we have just defined matches with the restricted Cartan operator $\dCh$ used in Sec.\ \ref{s3}, see Eq.\ \eqref{dChonPhi}--\eqref{dChonbarLai}. This is a basic ingredient of the Cartan model of equivariant cohomology. The relations \eqref{deltaact2} define the notion of equivariant trivial pair introduced in Sec.\ \ref{s3}. This notion can be naturally understood from the action of the Kalkman automorphism, which allows to make the link between the Weil and Cartan models of equivariant cohomology. This is discussed in particular at the end of Sec.\ \ref{cohosec}.

Equipped with the equivariant differential $\delta$, we can rewrite the total action \eqref{Stotelem} in the form
\be\label{Stotwithdelta} S_{\text{tot}} = S(\varphi,\phi) - \delta\psi_{G/H}\, ,\ee
for a partial gauge-fixing fermion
\be\label{psiGHelem} \psi_{G/H} = \bar\omega_{i}\Bigl(F_{i}(\varphi,\phi) - \frac{\xi}{2}\bar\La_{i}\Bigr) + \frac{\alpha}{2} f_{ijk}\bar\omega_{i}\bar\omega_{j}\omega_{k}\, ,\ee
a formula mimicking Eq.\ \eqref{psigen}. We could then go on and show, in the same spirit as for the usual BRST formalism, that actually any total action of the form \eqref{Stotwithdelta}, with a $H$-invariant gauge-fixing fermion of ghost number $\gh\psi_{G/H} = -1$, is suitable to define an action $S_{H}(\varphi;\psi_{G/H})$ by the formula \eqref{SHelemdef}.
It is straightforward to check, along the lines explained previously, that all such actions $S_{H}$ define a unique partition function \eqref{ZHdef} which coincides with the partition function \eqref{Zdef} of the original, $G$-invariant gauge theory. The main result of Sec.\ \ref{s3} is then reproduced. 

\subsection{The path integral approach}

Let us now finally rederive once again some of our results, but this time from a reasoning \`a la Faddeev-Popov \cite{FP} in the context of the path integral. This will shed an interesting new light on some aspects of the formalism, most notably on the quartic ghost terms that are crucially needed in the partial gauge fixing.

\subsubsection*{Textbook full gauge-fixing in the path integral}

To put things into perspective, let us briefly review the usual textbook approach. One starts from the ill-defined path integral
\be\label{Zill} Z_{\text{ill-defined}} = \int\!D\Phi\, e^{-S(\Phi)}\, ,\ee
which is infinite because of the gauge invariance of the action $S$. The trick to obtain a well-defined path integral is to factorize the volume of the gauge group. To do so, one considers a set of gauge-fixing conditions $F_{A}$ and arbitrary external background fields $b_{A}$ and sets
\be\label{Deltafunc} \frac{1}{\Delta (\Phi,b)} = \int\!D\mu_{G}(g) \,
\delta\bigl(F_{A}(g\cdot\Phi) - b_{A}\bigr)\, ,\ee
where $D\mu_{G}(g)$ denotes the Haar measure on the gauge group and $g\cdot\Phi$ the action of the gauge transformation $g\in G$ on the physical fields. From the invariance of the Haar measure under right multiplication, $\Delta(\Phi,b)$ is automatically gauge invariant. If we note $g_{\Phi,b}$ the unique solution to the equations $F_{A}(g\cdot\Phi)=b_{A}$, we thus have
\be\label{Deltacomp}\Delta(\Phi,b)=\Delta(g_{\Phi,b}\cdot\Phi,b) = \det\L_{B}F_{A}(g_{\Phi,b}\cdot\Phi)\, ,\ee
where the last equality comes from the fact that the corresponding integral \eqref{Deltafunc} over the gauge group is dominated by the infinitesimal neighbourhood of the identity (we assume that there is no Gribov ambiguity and that the determinant on the right-hand side of \eqref{Deltacomp} is positive, as usual). One then writes formally
\begin{align}\label{alaFP0} \int\!D\Phi\, e^{-S(\Phi)} &= 
\int\!D\Phi\, e^{-S(\Phi)}\biggl(\Delta(\Phi,b) \int\!D\mu_{G}(g) \,
\delta\bigl(F_{A}(g\cdot\Phi) - b_{A}\bigr)\biggr)\\\label{alaFP1}
& = \int\! D\mu_{G}(g) \, I(g,b)\, ,
\end{align}
where
\be\label{Iintdef} I(g,b) = \int\! D\Phi\, e^{-S(\Phi)}\Delta(\Phi,b) \delta\bigl(F_{A}(g\cdot\Phi) - b_{A}\bigr) \, .\ee
In the first line of \eqref{alaFP0}, one has simply inserted 1 in the integrand of \eqref{Zill}, and in the second line one has formally inverted the integrals over $\Phi$ and over the gauge group. Now, using the gauge invariance of the measure $D\Phi e^{-S(\Phi)}$ and of $\Delta(\Phi,b)$, it is clear that $I(g,b)$ actually does not depend on $g$ (simply make the change of variable $\Phi\mapsto g^{-1}\cdot\Phi$),
\be\label{Inotg} I(g,b) = I(b)\, .\ee
This has two useful consequences. First, the integral over the gauge group factorizes in \eqref{alaFP1}, as looked for,
\be\label{alaFP2} Z_{\text{ill-defined}}= (\vol G) I\, .\ee
This also shows that $I(b)=I$ cannot depend on $b$.
Second, we can compute $I(g,b)$ by choosing $g=\text{Id}$ in \eqref{Iintdef}. Using \eqref{Deltacomp}, this yields
\be\label{alaFP3} I = \int\! D\Phi\, e^{-S(\Phi)}\delta\bigl(F_{A}(\Phi) - b_{A}\bigr) \det \L_{B}F_{A}(g_{\Phi,b}\cdot\Phi)\, .\ee
The $\delta$-function insertion in \eqref{alaFP3} ensures that the only field configurations that can contribute to the path integral are such that $g_{\Phi,b}\cdot\Phi = \Phi$. We thus get 
\be\label{alaFP4} Z = \frac{Z_{\text{ill-defined}}}{\vol G}
= I = \int\! D\Phi\, e^{-S(\Phi)} \delta\bigl(F_{A}(\Phi) - b_{A}\bigr)
\det \L_{B}F_{A}(\Phi) \, .\ee
The determinant is dealt with in the usual way by introducing the anticommuting ghosts and antighosts with the standard quadratic Faddeev-Popov Lagrangian,
\be\label{alaFP5} \det\L_{B}F_{A}(\Phi) = \int\! D\omega_{A}D\bar\omega_{A}
\, e^{-\bar\omega_{A}\L_{B}F_{A}\omega_{B}}\, .\ee
The interest in having introduced the arbitrary background fields $b_{A}$ is now apparent. Since the partition function \eqref{alaFP4} does not depend on them, we can treat them as Gaussian random variables and thus get rid of the strict $\delta$-function gauge-fixing conditions. Using for example a Gaussian weight of the form
\be\label{Gausswei} e^{-\frac{b_{a}b_{a}}{2\zeta} - \frac{b_{i}b_{i}}{2\xi}}\, ,\ee
we obtain
\be\label{alaFP6} Z = \int\! D\Phi D\omega_{A}D\bar\omega_{A}
\, e^{-S -\frac{F_{a}F_{a}}{2\zeta} - \frac{F_{i}F_{i}}{2\xi}
-\bar\omega_{A}\L_{B}F_{A}\omega_{B}}\, .\ee
We can then trivially integrate in the auxiliary fields $\bar\la_{a}$ and $\bar\la_{i}$ to put the partition function in the form \eqref{Zdef}, with $s_{G}\Psi_{G}$ being given by \eqref{sGonpsistan}.

\subsubsection*{Strict partial gauge-fixing in the path integral}

Let us now try to reproduce the above steps, but for a partial gauge-fixing of the gauge symmetry $G$ down to the non-trivial subgroup $H$. We thus use partial gauge-fixing conditions $F_{i}(\Phi)$ which preserve $H$, by satisfying the fundamental constraints \eqref{condonFibis}. In order to insert the partial gauge-fixing conditions into the path integral, we would like to define the analogue of $\Delta(\Phi,b)$, mimicking Eq.\ \eqref{Deltafunc},
\be\label{DelHdef} \frac{1}{\Delta_{H}(\Phi,b)} \overset{?}{=}
\int\! D\mu_{G}(g)\,\delta\bigl(F_{i}(g\cdot\Phi) - b_{i}\bigr)\, .\ee
With such a definition, $\Delta_{H}(\Phi,b)$ is automatically $G$-invariant, for any fixed background fields $b_{i}$. However, \emph{the background fields $b_{i}$ break the $H$ covariance of the gauge-fixing conditions}. In particular the volume of the group $H$ will not factorize in \eqref{DelHdef}, as one would have expected for a correct implementation of partial gauge-fixing conditions preserving $H$. This clash between the requirement of maintaining the gauge invariance under $H$ and the possibility to introduce the arbitrary background fields $b_{i}$ in the gauge-fixing conditions is the path integral version of the difficulty encountered before in Sec.\ \ref{diagosec} in the case $\xi\not = 0$. If we cannot use the background field trick, we will not be able to get a Gaussian weight for the gauge-fixing conditions as in \eqref{alaFP6}.

Before explaining how to go around this difficulty in the path integral framework (which we know from our previous discussion can be solved in principle by introducing quartic ghost couplings), let us abandon the background field $b$ and work with 
\be\label{DelHdef2} \frac{1}{\Delta_{H}(\Phi)} =
\int\! D\mu_{G}(g)\,\delta\bigl(F_{i}(g\cdot\Phi)\bigr)\, .\ee
Thanks to the covariance \eqref{condonFibis} of $F_{i}$ with respect to $H$, the general solution to the equations $F_{i}(g\cdot\Phi)=0$ is of the form $h g_{\Phi}$, for some gauge transformation $g_{\Phi}$ and any gauge transformation $h\in H$. The $\delta$-function in \eqref{DelHdef2} thus leaves free an integral over $H$ and the volume of $H$ factorizes, yielding
\be\label{DelHsol} \Delta_{H}(\Phi) = \frac{1}{\vol H}
\det\L_{j}F_{i}(g_{\Phi}\cdot\Phi)\, .\ee
This formula is consistent in spite of the ambiguity in the definition of $g_{\Phi}$ by left multiplication by an arbitrary element of $H$, because the matrix $\L_{j}F_{i}$ transforms in the real unitary representation $R_{G/H}\otimes R_{G/H}$ of $H$ and the determinant is invariant under such a transformation. Again, we have assumed that there is no Gribov ambiguity and that the determinant on the right-hand side of \eqref{DelHsol} is positive. Formally inserting 1 in \eqref{Zill} by using \eqref{DelHdef2}, we can then proceed exactly as in the standard Faddeev-Popov derivation, from Eq.\ \eqref{alaFP0} to \eqref{alaFP5}. Implementing the $\delta$-function constraints with a Lagrange multuplier $\bar\La_{i}$, we get in this way
\be\label{newFP} Z = \frac{1}{\vol G}\int\!D\Phi\, e^{-S(\Phi)} =
\frac{1}{\vol H}\int\! D\Phi D\omega_{i}D\bar\omega_{i}D\bar\La_{i}\, 
e^{-S(\Phi) - \bar\La_{i}F_{i}-\bar\omega_{i}\L_{j}F_{i}\omega_{j}}\, .\ee
We have thus found again the correct partial gauge-fixing formula in the case of gauge-fixing conditions imposed strictly, which corresponds to the total action \eqref{Stotelem} with $\xi=\alpha=0$.

\subsubsection*{Partial gauge-fixing with a Gaussian weight in the path integral}

How can we implement the partial gauge-fixing conditions with a Gaussian weight in the path integral? Of course, we could now use the equivariant cohomological method. Since we have just proven that the formula \eqref{Stotwithdelta} is correct for the particular gauge-fixing fermion \eqref{psiGHelem} at $\xi=\alpha=0$, we could show directly using the standard path integral argument based on $\delta S(\Phi)=0$ and \eqref{deltasq} that it is also correct for any $H$-invariant gauge-fixing fermion, in particular for \eqref{psiGHelem} at any values of $\xi$ and $\alpha$. But we rather seek here a direct path integral argument \`a la Faddeev-Popov that would automatically produce a Gaussian weight for the conditions $F_{i}$ and the required quartic terms in the ghost Lagrangian.

Since we cannot use the familiar trick of introducing external background fields, as explained in the previous paragraph, the idea is to introduce instead new \emph{dynamical} fields. A trick of this kind was first used by Zinn-Justin in \cite{ZJtrick}. Of course, the dynamics of these fields must not change the partition function $Z$ and thus must be essentially trivial. We shall simply consider an adjoint scalar field $h_{A}$ and modify the original gauge theory by including a $G$-invariant quadratic term in $h$,
\be\label{Smodif} S(\Phi)\rightarrow S(\Phi) + \frac{h_{A}h_{A}}{2\zeta}\,\cdotp\ee
Let us emphasize that it is crucial for the modification of the action to be $G$-invariant. In this case one can prove straightforwardly, using a gauge-fixing condition that does not depend on $h$, that the inclusion of $h$ does not change $Z$.

Let us now consider \emph{strict} partial gauge fixing conditions of the form
\be\label{stricth} F_{i} = \mathcal F_{i} - h_{i}=0\, ,\ee
where the $\mathcal F_{i}$ do not depend on $h$. Since these are strict conditions, we can use the formula \eqref{newFP} derived in the previous paragraph without change if not for \eqref{Smodif}. The crucial point is that since the $h_{i}$ are dynamical quantum fields, not classical background fields, the gauge transformations act non-trivially on them and we have
\be\label{LjFi} \L_{j}F_{i} = \L_{j}\mathcal F_{i} + f_{ijk}h_{k}+f_{ija}h_{a}\, .\ee
The integrals over $h_{a}$ and $h_{i}$ can then be done straightforwardly in \eqref{newFP}, taking into account the quadratic $h$-terms in \eqref{Smodif} and the linear $h$-terms in \eqref{LjFi}. These Gaussian path integrals automatically produce a Gaussian weight for $\mathcal F_{i}$ and quartic ghost couplings! Actually, one can easily check, using the Jacobi identity, that the resulting formula precisely matches the total action \eqref{Stotelem}, with $F_{i}$ replaced by $\mathcal F_{i}$ and $\zeta=\xi=\alpha$.

\noindent\emph{Remark}: the reader might wonder what happens if we use the above procedure in \eqref{psiGHelem}, chosing $F_{i}$ as in \eqref{stricth} and integrating out $h_{A}$ at the end. The result is that one goes back to a total action of the form \eqref{Stotelem}, with $F_{i}$ replaced by $\mathcal F_{i}$ and the parameters $\xi$ and $\alpha$ both shifted by $\zeta$. 

\section{Example: general Yang-Mills D-brane systems} \label{s5}

As a simple application of our general formulas, let us present the case of the pure (Euclidean, four dimensional) Yang-Mills theory with 
\be\label{GHex} G=\text{U}(N)\, ,\quad H=\text{U}(N_{1})\times\cdots\times\text{U}(N_{r})\, .\ee
Let us note that it is slightly more convenient to use $G=\uN$ instead of $G=\text{SU}(N)$, but the two cases are essentially equivalent since the global $\text{U}(1)$ in the $\uN$ theory decouples. As explained in \cite{ferfund}, this example represents the general pure Yang-Mills D-brane system, where we consider $r$ stacks of branes each built from $N_{I}$ branes, with
\be\label{NNI} N = \sum_{I=1}^{r}N_{I}\, .\ee
When a subset of the $\{N_{I}\}$ are much larger than the others, it is natural to gauge fix further down by eliminating the associated $\text{U}(N_{I})$ factors and integrating out the associated fields. The resulting model then corresponds to probe branes moving in the emergent holographic geometry sourced by the branes that have been integrated out \cite{ferfund}. Another interesting special case is to take $r=N$ and thus set all the $N_{I}=1$. This is 't~Hooft Maximal Abelian Gauge.

\subsubsection*{Setting-up notations}

The only physical elementary field $\Phi$ in this example is the vector potential $\mathscr A_{\mu}$. As is usually done in the physics literature, we choose the $\mathscr A_{\mu}$ to be Hermitian matrices. They decompose as
\be\label{Amudec} \mathscr A_{\mu} = i\mathscr A_{\mu A}\tau_{A} \ee
on our usual basis $(\tau_{A})$ of the Lie algebra $\mathfrak g = \mathfrak u(N)$. We also have the ghosts $\omega$ and the equivariant trivial pair ($\bar\omega,\bar\La)$, which decompose on the broken generators of $\mathfrak g/\mathfrak h$ only,
\be\label{ghostdec} \omega = i\omega_{i}\tau_{i}\, ,\quad
\bar\omega = i\bar\omega_{i}\tau_{i}\, ,\quad \bar\La = i\bar\La_{i}\tau_{i}\, .\ee

We decompose any Hermitian matrix $M$ in block matrices $M^{IJ}$ of sizes $N_{I}\times N_{J}$, corresponding to the various unbroken $\text{U}(N_{I})$ factors. Explicitly, in terms of matrix components,
\be\label{defAmublocks} \bigl(M^{IJ} \bigr)_{\alpha\beta} = 
M_{\sum_{K<I}N_{K}+\alpha,\,  \sum_{K<J}N_{K}+\beta}\, ,\quad 1\leq \alpha\leq N_{I}\, ,\ 1\leq\beta\leq N_{J}\, .
\ee
In particular,
\be\label{hermconj} \bigl(M^{IJ}\bigr)^{\dagger} = M^{JI}\, .\ee
If we write
\be\label{Montaudec} M = iM_{A}\tau_{A} = iM_{a}\tau_{a} + i M_{i}\tau_{i}\, ,\ee
the ``diagonal'' blocks $M^{II}$ correspond to the ``unbroken'' $\mathfrak h$-indices $a$ and the ``off-diagonal'' blocks $M^{IJ}$ for $I\not = J$ correspond to the ``broken'' $\mathfrak g/\mathfrak h$-indices. For the vector potential, it is convenient to introduce the notations
\be\label{AmuWmu} \mathscr A_{\mu}^{II}  = A_{\mu}^{I}\, ,\quad \mathscr A_{\mu}^{IJ}  = W_{\mu}^{IJ}\ \text{for} \ I\not = J\, .\ee
For the other variables $\omega$, $\bar\omega$ and $\bar\La$, only the off-diagonal blocks are present and thus
\be\label{diagvanish} \omega^{II} = \bar\omega^{II}=\bar\La^{II}=0\, .\ee
The trace over the $N_{I}\times N_{I}$ diagonal blocks will be denoted by $\tr_{I}$. So, for example, we can consider $\tr_{I}M^{II}$ (no sum over $I$!).

One last convention we are going to use to get rid of any ambiguity in our formulas is that sums over indices of the type $I,J,K,\ldots,$ labeling the various block matrices defined above will always be indicated explicitly with a sum sign. In particular, no sum is a priori implied if two such indices are repeated in an expression.  An example of this convention is the definition of the $H$-covariant derivative $\nabla_{\mu}$, which acts as
\be\label{nablaHdef} \nabla_{\mu}M^{IJ} = \partial_{\mu}M^{IJ} + i A_{\mu}^{I}M^{IJ} - i M^{IJ}A_{\mu}^{J}\, .\ee
Of course, for all the other types of indices, the usual Einstein summation convention is kept.

All we have to do now is to straightforwardly apply the formulas of the previous sections, using the above notations which are specially adapted to the case \eqref{GHex}.

\subsubsection*{The general case}

\noindent\emph{The equivariant differential}

Equations \eqref{dChonPhi}--\eqref{dChonbarLai}, or equivalently \eqref{deltaact1} and \eqref{deltaact2}, yield
\begin{align}\label{dex1} &\delta A_{\mu}^{I} = i\sum_{J}\bigl(
\omega^{IJ}W_{\mu}^{JI} - W_{\mu}^{IJ}\omega^{JI}\bigr)\\
\label{dex2} &\delta W_{\mu}^{IJ} =  - \nabla_{\mu}\omega^{IJ} + i
\sum_{K}\bigl(\omega^{IK}W_{\mu}^{KJ} - W_{\mu}^{IK}\omega^{KJ}\bigr)\quad (I\not = J)\\
\label{dex3}& \delta\omega^{IJ} = i\sum_{K}\omega^{IK}\omega^{KJ}\quad
(I\not = J)\\\label{dex4}
& \delta\bar\omega^{IJ} = -\bar\La^{IJ}\, ,\quad
 \delta\bar\La^{IJ} = \sum_{K}\bigl(\bar\omega^{IJ}\omega^{JK}\omega^{KJ} - \omega^{IK}\omega^{KI}\bar\omega^{IK}\bigr)\, .
\end{align}
The curvature is a matrix $R$ with only diagonal blocks given by
\be\label{RIex} R^{I} = i\sum_{J}\omega^{IJ}\omega^{JI}\, .\ee
In particular
\be\label{delta2ex} \delta^{2} = -\L_{R}\, ,\ee
where $\L_{\varepsilon}$ generates $\text{U}(N_{1})\times\cdots\times\text{U}(N_{r})$ gauge transformations with gauge parameters $\varepsilon^{I}$,
\be\label{gHex1} \L_{\varepsilon}A_{\mu}^{I} = -\nabla_{\mu}\varepsilon^{I}\, ,\quad \L_{\varepsilon}M^{IJ} = i\bigl(\varepsilon^{I}M^{IJ} - M^{IJ}\varepsilon^{J}\bigr)\, .\ee
The matrices $M^{IJ}$ in the above equation could be $W_{\mu}^{IJ}$, $\omega^{IJ}$, $\bar\omega^{IJ}$ or $\bar\La^{IJ}$.

\medskip

\noindent\emph{The gauge-fixing fermion}

The gauge-fixing fermion \eqref{psigen} or \eqref{psiGHelem} reads
\be\label{psiGHex} \psi_{G/H} = \int\! \d x\,\sum_{I,J}\tr_{I}\bar\omega^{IJ}
\Bigl( F^{JI} - \frac{\xi}{2}\bar\La^{JI} - i\alpha\sum_{K}\bar\omega^{JK}
\omega^{KI}\Bigr)\, ,\ee
for gauge-fixing conditions
\be\label{FIJ} F^{IJ} = \nabla_{\mu}^{}W_{\mu}^{IJ} = \partial_{\mu}^{}W_{\mu}^{IJ} + i A_{\mu}^{I}W_{\mu}^{IJ} - iW_{\mu}^{IJ}A_{\mu}^{J}\, ,\quad I\not = J\, .\ee

\medskip

\noindent\emph{The partial gauge-fixing plus ghost terms}

One can then compute 
\begin{multline}\label{totSex} - \delta\psi_{G/H} = \int\!\d x\,\sum_{I,J}\tr_{I}\Biggl[ \bar\La^{IJ}\Bigl( F^{JI} - \frac{\xi}{2}\bar\La^{JI}
-i\alpha\sum_{K}\bigl(\bar\omega^{JK}\omega^{KI} + \omega^{JK}\bar\omega^{KI}\bigr)\Bigr)\\
+ \nabla_{\mu}\bar\omega^{IJ}\nabla_{\mu}\omega^{JI} + i\nabla_{\mu}\bar\omega^{IJ}\sum_{K}\bigl(W_{\mu}^{JK}\omega^{KI}-\omega^{JK}W_{\mu}^{KI}\bigr)\\
+ \sum_{K}\Bigl(\bigl(\omega^{IJ}\bar\omega^{JK}-\bar\omega^{IJ}\omega^{JK}\bigr)W_{\mu}^{KJ}W_{\mu}^{JI} + \bar\omega^{IJ}\bigl(
W_{\mu}^{JK}\omega^{KJ}W_{\mu}^{JI}+W_{\mu}^{JI}\omega^{IK}W_{\mu}^{KI}\bigr)\Bigr)\\
-\xi\sum_{K}\bar\omega^{IJ}\bar\omega^{JI}\omega^{IK}\omega^{KI} - 
\alpha\sum_{K\not = I, L}\bar\omega^{IJ}\bar\omega^{JK}\omega^{KL}\omega^{LI}\Biggr]\, ,
\end{multline}
consistently with \eqref{dChonpsi} or \eqref{Stotelem}.

Let us note that the formulas can be simplified a bit in the maximal Abelian projection case $\text{U}(N)\rightarrow\text{U}(1)^{N}$, since then the variables are no longer matrices but either commuting or anticommuting numbers.

\subsubsection*{The symmetric space case}

At rank $r=2$, the above formulas greatly simplify, because then $f_{ijk}=0$ and the quotient $G/H$ is a symmetric space. In particular, the parameter $\alpha$ does not enter. Since we then have only two types of off-diagonal blocks, $M^{12}$ and $M^{21}$, it is convenient to note
\be\label{Mpmdef} M^{-} = M^{12}\, ,\quad M^{+} = M^{21} = \bigl(M^{-}\bigr)^{\dagger}\, .\ee
Equations \eqref{dex1}--\eqref{totSex} then simplify to
\begin{align}
\label{f1} &\delta A_{\mu}^{1} = i\bigl(\omega^{-}W_{\mu}^{+} - 
W_{\mu}^{-}\omega^{+}\bigr)\, ,\quad 
\delta A_{\mu}^{2} = i\bigl(\omega^{+}W_{\mu}^{-} - 
W_{\mu}^{+}\omega^{-}\bigr)\\ 
\label{f2}& \delta W_{\mu}^{\mp} = -\nabla_{\mu}\omega^{\mp}\\
\label{f3} &\delta\omega^{\mp} = 0\, ,\quad \delta\bar\omega^{\mp} = -\bar\La^{\mp}\\\label{f4}&
\delta\bar\La^{\mp} =\bar\omega^{\mp}\omega^{\pm}\omega^{\mp} -\omega^{\mp}
\omega^{\pm}\bar\omega^{\mp}\, ,
\end{align}
\begin{gather}\label{d2d2d2} \delta^{2} = -\L_{R}\\\label{fRex} R^{1} = i\omega^{-}\omega^{+}\, ,\quad R^{2} = i\omega^{+}\omega^{-}\, ,
\end{gather}
\begin{gather}\label{gHsymmex} \L_{\varepsilon}A_{\mu}^{1} = -\nabla_{\mu}\varepsilon^{1}\, ,\quad \L_{\varepsilon}A_{\mu}^{2} = -\nabla_{\mu}\varepsilon^{2}\\\label{gHsymmex2} \L_{\varepsilon}M^{-} = i\bigl(\varepsilon^{1}
M^{-}-M^{-}\varepsilon^{2}\bigr)\, ,\quad \L_{\varepsilon}M^{+} = i\bigl(\varepsilon^{2}M^{+}-M^{+}\varepsilon^{1}\bigr)\, ,
\end{gather}
\begin{align}\label{f5}&
\psi_{G/H} = \int\!\d x\, \biggl[ \tr_{1}\bar\omega^{-}\Bigl(
F^{+}- \frac{\xi}{2}\bar\La^{+}\Bigr) + \tr_{2}\bar\omega^{+}\Bigl(
F^{-}- \frac{\xi}{2}\bar\La^{-}\Bigr)\biggr]
\\\label{f6}&
F^{\mp} = \nabla_{\mu}^{}W_{\mu}^{\mp}\, ,
\end{align}
\begin{multline}
-\delta\psi_{G/H} = \int\!\d x\,\biggl[\tr_{1}\Bigl[\bar\La^{-}\Bigl(F^{+} -
\frac{\xi}{2}\bar\La^{+}\Bigr) + \nabla_{\mu}\bar\omega^{-}\nabla_{\mu}
\omega^{+}\\ + \bigl(\omega^{-}\bar\omega^{+}-\bar\omega^{-}\omega^{+}\bigr)W_{\mu}^{-}W_{\mu}^{+} + 2 \bar\omega^{-}W_{\mu}^{+}\omega^{-}W_{\mu}^{+} -\xi\bar\omega^{-}\bar\omega^{+}\omega^{-}\omega^{+}\Bigr]
+ \tr_{2}\Bigl[\cdots\Bigr]\biggr]\, ,
\end{multline}
where the $\cdots$ in the $\tr_{2}\bigl[\cdots\bigr]$ terms are exactly the same as in the $\tr_{1}\bigl[\cdots\bigr]$ terms, except that all the $\mp$ signs must be flipped to $\pm$. The equations simplify even further in the much studied case $\text{U}(2)\rightarrow\text{U}(1)^{2}$, since then the variables are no longer matrices but simply commuting or anticommuting numbers.

\vfill\eject

\section{\label{s6} Conclusion}

We have provided a full solution to the problem of partially fixing the gauge from a gauge group $G$ down to a non-trivial subgroup $H\subset G$ in standard gauge theories of Yang-Mills type, as outlined in Sec.\ \ref{problemsec}, using different points of view and methods. The partial gauge fixing is governed by an equivariant BRST cohomology, yielding many different $H$-invariant actions $S_{H}$ which are all strictly equivalent to the original $G$-invariant model. A crucial feature of the partial gauge-fixing procedure is that the ghost action includes quartic ghost terms, even at tree-level. These quartic ghost terms are essential to ensure the gauge invariance of the model and in particular the fact that all the actions $S_{H}$, which strongly depend on the partial gauge-fixing conditions one uses, all yield the same gauge-invariant partition function.

Several extensions and generalizations of our work seem possible and worth pursuing. First, it is likely that an equivariant anti-BRST symmetry associated with the partial gauge-fixing exists, even though no sign of it seems to be present in the standard models of equivariant cohomology. We leave this construction, which would complement nicely the formalism we have developed, for a future work. Second, it would be nice to explicitly generalize our formulas to the case of reducible gauge symmetries and, even more interestingly, to open gauge algebras. A comprehensive theory should look like some equivariant version of the BV formalism. Third, it is very desirable to build a supersymmetric version of the partial gauge-fixing formalism, for minimal and extended versions of supersymmetry, which could be used in the context of supersymmetric gauge theories. The non-trivial goal is to define supersymmetric versions of our actions $S_{H}$. This is particularly important in view of applications to D-brane systems, since supersymmetric D-brane configurations have been much studied and have remarkable properties. Finally, our analysis has been limited to tree-level and a detailed study of the consequences of the equivariant BRST symmetry at the quantum level is required. In this respect, the results in \cite{renAG} seem very encouraging.

Our prime motivation for the present work was to develop the required tools to build holographic emergent space models based on the microscopic description of probe D-branes \cite{fer1,ferfund}. It was by working out these tools that we realized that there are deep and unexpected connections between the D-brane story, the Abelian projection scenario and the effective low energy description of gauge theories for which the gauge group $G$ is Higgsed down to $H$. As explained in \cite{ferfund}, the holographic emergent dimensions are the analogue of the gluon and ghost/antighost condensates considered in the literature on the Abelian projection and much studied, in particular, in lattice gauge theory (see e.g.\ \cite{Kondo,Schaden} and references therein). Moreover, the gauge-dependence of the constructions, which, in our formalism, shows up through the dependence of the actions $S_{H}$ on the partial gauge-fixing fermion $\psi_{G/H}$, is then related to the problem of bulk locality in quantum gravity. This problem is itself mapped to the usual $H$-invariant, but not $G$-invariant, description of the low-energy physics in grand unified models in which a grand unified gauge group $G$ is broken down to $H$. We hope that these new ideas will open the way for many interesting applications of our results.

\subsubsection*{Acknowledgments}

This work is supported in part by the Belgian FRFC (grant 2.4655.07) and IISN (grants 4.4511.06 and 4.4514.08).

%
%

%
%

%

%

%
\end{document}